\documentclass[10pt]{iopart}
\usepackage{iopams}
\usepackage{graphicx}
\usepackage{mathrsfs}
\begin{document}

\title[]{A novel scheme for modelling dissipation (gain) and thermalization in open quantum systems}

\author{F. Kheirandish, E. Bolandhemmat, N. Cheraghpour, R. Moradi and S. Ahmadian}

\address{Department of Physics, University of Kurdistan, P.O.Box 66177-15175, Sanandaj, Iran}
\ead{f.kheirandish@uok.ac.ir}
\vspace{10pt}
\newcommand{\bra}[1]{\langle #1 \vert}
\newcommand{\ave}[1]{\langle #1 \rangle}
\newcommand{\ket}[1]{\vert #1 \rangle}
\newcommand{\la}{\langle}
\newcommand{\ra}{\rangle}
\newcommand{\commentold}[1]{}
\newcommand{\ah}{\hat{a}}
\newcommand{\ahd}{\hat{a}^\dag}
\newcommand{\bh}{\hat{b}}
\newcommand{\bhd}{\hat{b}^\dag}
\newcommand{\ch}{\hat{c}}
\newcommand{\chd}{\hat{c}^\dag}
\newcommand{\Ah}{\hat{A}}
\newcommand{\Ahd}{\hat{A}^\dag}
\newcommand{\Bh}{\hat{B}}
\newcommand{\Bhd}{\hat{B}^\dag}
\newcommand{\roh}{\hat{\rho}}
\newcommand{\Hh}{\hat{H}}
\newcommand{\Uh}{\hat{U}}
\newcommand{\Uhd}{\hat{U}^\dag}
\newcommand{\ot}{\otimes}
\newcommand{\om}{\omega}
\newcommand{\Om}{\Omega}
\newcommand{\nn}{\nonumber}
\newcommand{\be}{\begin{equation}}
\newcommand{\ee}{\end{equation}}
\newcommand{\beq}{\begin{eqnarray}}
\newcommand{\eeq}{\end{eqnarray}}
\begin{abstract}
\noindent In this letter, we introduce a novel method for investigating dissipation (gain) and thermalization in an open quantum system. In this method, the quantum system is coupled linearly with a copy of itself or with another system described by a finite number of bosonic operators. The time-dependent coupling functions play a fundamental role in this scheme. To demonstrate the efficiency and significance of the method, we apply it to some ubiquitous open quantum systems. Firstly, we investigate a quantum oscillator in the presence of a thermal bath at the inverse temperature $\beta$, obtaining the reduced density matrix, the Husimi distribution function, and the quantum heat distribution function accurately. The results are consistent with existing literature by appropriate choices for the time-dependent coupling function. To illustrate the generalizability of this method to systems interacting with multiple thermal baths, we study the interaction of a quantum oscillator with two thermal baths at different temperatures and obtain compatible results. Subsequently, we analyze a two-level atom with energy or phase dissipation and derive the spontaneous emission and the pure dephasing processes consistently using the new method. Finally, we investigate the Markovianity in a dissipative two-level system.
\end{abstract}
\section{Introduction}
\noindent In the fascinating realm of quantum mechanics, understanding how systems interact with their surrounding environment is crucial for various phenomena, ranging from energy transfer processes to the stability of quantum states \cite{Feynman2010}. One of the fundamental concepts in this domain is quantum dissipation, which elucidates the dynamics of systems as they lose energy to their surroundings, leading to the gradual decay of their coherence and superposition states \cite{Breuer2002}.
At its core, quantum dissipation explores the intricate interplay between a quantum system of interest and its surrounding environment, often referred to as a bath \cite{Leggett1987}. This interaction introduces a myriad of complex phenomena that challenge traditional classical intuitions. Unlike classical dissipation, where energy loss is typically attributed to macroscopic friction or resistance, quantum dissipation delves into the probabilistic nature of quantum states and their evolution under the influence of environmental perturbations \cite{Weiss2012}.
A key aspect of quantum dissipation lies in the concept of decoherence, wherein the coherence and superposition of quantum states degrade over time due to interactions with the environment \cite{Zurek2003}. This process leads to the emergence of classical-like behavior, effectively erasing the delicate quantum features that distinguish quantum systems from their classical counterparts. Understanding and controlling decoherence is paramount for various applications in quantum information processing, quantum computing, and quantum communication \cite{Nielsen2010}.
Moreover, the study of quantum dissipation encompasses a rich theoretical framework, drawing from diverse fields such as quantum mechanics, statistical physics, and open quantum systems theory \cite{Breuer2002}. Researchers employ various mathematical tools, including master equations, stochastic methods, and quantum field theory, to model and analyze dissipation processes accurately \cite{Carmichael1999}.
Quantum dissipation finds relevance in a wide array of physical systems, ranging from microscopic quantum circuits and atoms to macroscopic condensed matter systems and cosmological phenomena \cite{Leggett1987}. Investigating dissipation mechanisms sheds light on phenomena like relaxation processes, energy transport, and the emergence of irreversible dynamics in quantum systems \cite{Weiss2012}.
Furthermore, quantum dissipation plays a crucial role in elucidating the behavior of quantum systems far from equilibrium. By probing the dynamics of dissipative quantum systems, researchers gain insights into the emergence of novel phenomena such as quantum phase transitions, nonequilibrium steady states, and quantum criticality \cite{Diehl2008}.
In conclusion, quantum dissipation stands as a cornerstone in the study of quantum phenomena, offering profound insights into the intricate dynamics of quantum systems interacting with their environment \cite{Breuer2002}. By unraveling the mysteries of dissipation, researchers pave the way for harnessing quantum coherence, controlling quantum dynamics, and unlocking the full potential of quantum technologies in the quest for understanding the quantum nature of reality.

Our goal in this article is introducing a novel method for investigating dissipation (gain) and thermalization in an open quantum system. For this purpose, the main quantum system is coupled linearly with a copy of itself or with another system described by a finite number of bosonic operators trough time-dependent coupling functions, which play a fundamental role in this scheme. Throughout the article, the results derived from this method are not only straightforward to obtain but are also consistent with those obtained from the Lindblad master equation, motivating further exploration of this idea. Here, we do not claim that the present method can serve as a replacement for the established approaches used to study dissipative quantum systems. Until then, we have a long and challenging journey ahead of us.
\section{Setting the stage}
\noindent Consider a quantum system with Hamiltonian $\hat{H}_a$ described by operators $\{\hat{a}_i, \hat{a}^\dag_i\}^N_{i=1}$, as the main system. To model loss or thermalization within this system, we linearly couple the main system to either a copy of itself or to a bath system with Hamiltonian $\hat{H}_b$, described by a finite set of bosonic operators $\{\hat{b}_j, \hat{b}^\dag_j\}^N_{j=1}$.
The coupling functions are assumed time-dependent. The total Hamiltonian is given by
\be\label{H}
\hat{H}=\hat{H}_a+\hat{H}_b+\sum_{i,j} g_{ij} (t)[\hat{a}_i \hat{b}^\dag_j+\hat{a}^\dag_i \hat{b}_j],
\ee
where the time-dependent coupling functions $g_{ij} (t)$ play a crucial role in the introduced loss or gain scheme. The commutation relations among these operators are
\beq\label{Commutations}
&[\ah_i, \ahd_j]=\delta_{ij},\nn\\
&[\bh_i, \bhd_j]=\delta_{ij},\nn\\
&[\ah_i, \bh_j]=[\ah_i, \bhd_j]=0.
\eeq
We need to find the time-evolution of operators $\ah_i (t)$ and $\bh_j (t)$ in Heisenberg picture
\beq\label{Hab}
&&\dot{\ah}_i (t)=-\frac{i}{\hbar}[\ah_i (t), \hat{H}_a]-\frac{i}{\hbar}\sum_{j=1}^N g_{ij} (t)\,\bh_j (t),\nn\\
&&\dot{\bh}_i (t)=-\frac{i}{\hbar}[\bh_i (t), \hat{H}_b]-\frac{i}{\hbar}\sum_{j=1}^N g_{ji} (t)\,\ah_j (t).
\eeq
Equations Eq.~(\ref{Hab}) may have explicit analytical solutions, otherwise, we invoke approximate solutions. By selecting decreasing coupling functions over time, we demonstrate that the energy flowing from the main system to the bath system cannot completely return to the main system.
\section{Dissipative harmonic oscillator}
\begin{figure}[t]
    \centering
    \includegraphics[width=.4\textwidth]{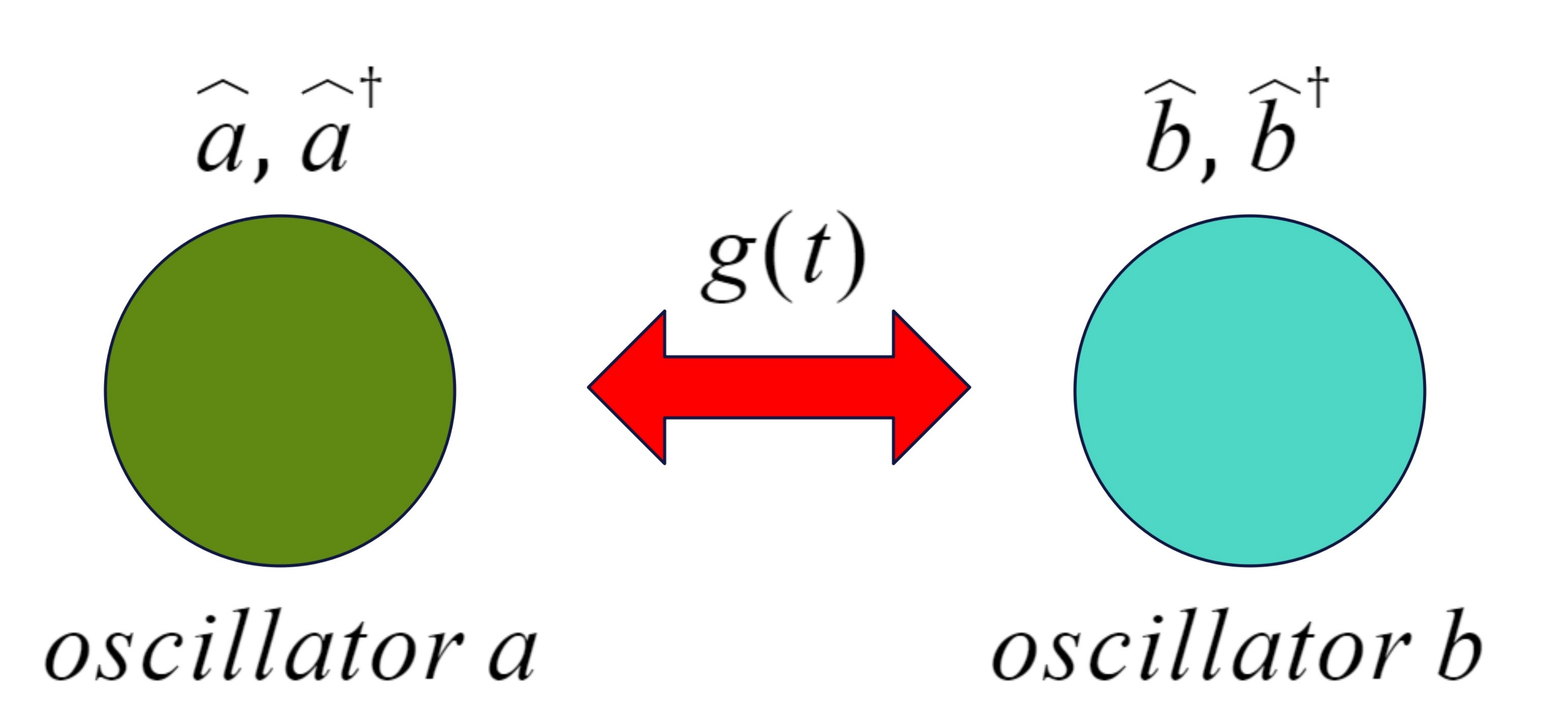}
    \caption{(Color online) Two interacting oscillators with a time dependent coupling function $g(t)$.}\label{figure1}
\end{figure}

\noindent Let us embark on our journey to demonstrate the effectiveness of the scheme. As our initial example, let's consider the ubiquitous quantum harmonic oscillator interacting with a heat bath characterized by the inverse temperature $\beta=1/\kappa_B T$, where $\kappa_B$ is the Boltzmann's constant \cite{Dekker1981,Serhan2018,Kaur2021}. The bath oscillator mirrors the main oscillator, described by the ladder operators $\hat{b},\,\hat{b}^\dag$. These two oscillators interact through a time-dependent coupling function $g(t)$, see Fig.~(\ref{figure1}).
The Hamiltonian is given by
\be\label{H1}
\Hh(t)=\hbar\om_0\,\ahd\ah+\hbar\om_0\bhd\bh+\hbar g(t)\,(\ah\bhd+\ahd\bh).
\ee
Using the Bogoliubov transformations
\beq\label{Bogo}
&& \ah=\frac{1}{\sqrt{2}}(\Ah+\Bh),\nn\\
&& \bh=\frac{1}{\sqrt{2}}(\Bh-\Ah),
\eeq
the Hamiltonian becomes separable in terms of the new ladder operators
\be\label{HAB}
\Hh(t)=\hbar\om_A (t)\,\Ahd\Ah+\hbar\om_B (t)\,\Bhd\Bh,
\ee
where we have defined $\om_A (t)=\om_0-g(t)$, $\om_B (t)=\om_0+g(t)$. The time-evolution operator corresponding to Hamiltonian (\ref{H1}) is
\be\label{U1}
\Uh (t)=e^{-i\Om_A (t)\,\Ahd\Ah}\ot e^{-i\Om_B (t)\,\Bhd\Bh},
\ee
where $\Om_{A(B)} (t)=\int_0^t dt'\,\om_{A(B)} (t')$.

\begin{figure}[t]
    \centering
    \includegraphics[width=.4\textwidth]{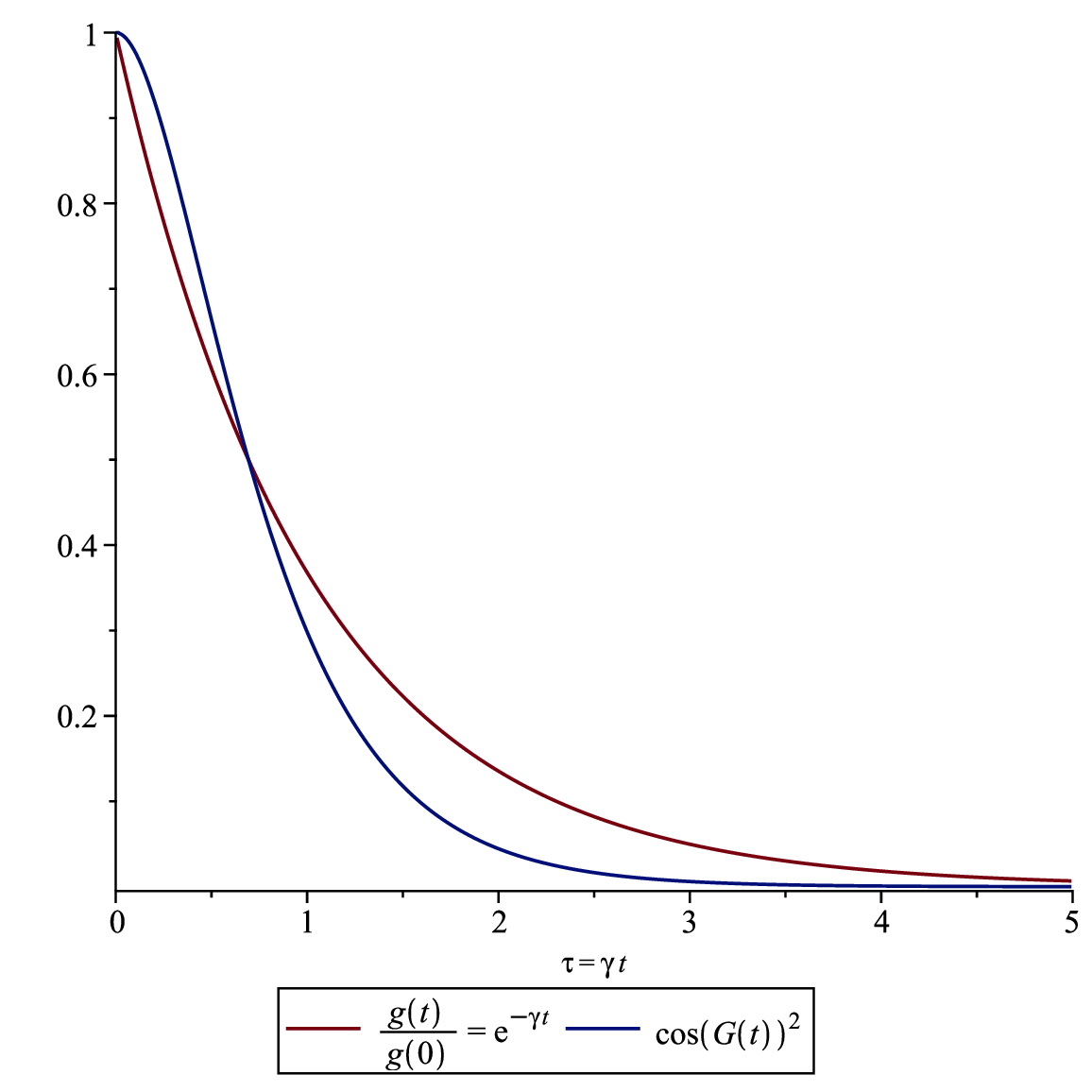}
    \caption{(Color online) The nearly identical behavior of functions $g(t)/g(0)=e^{-\gamma t}$ and $\cos^{2} G(t)=\sin^2(\frac{\pi}{2}e^{-\gamma t})$ in terms of $\tau=\gamma t$.}\label{comparison}
\end{figure}

The time evolution of the total density matrix is described by
\be\label{R1}
\roh(t)=\Uh(t)\roh(0)\Uhd,
\ee
where $\roh(0)$ is typically chosen as a separable state $\roh_a (0)\ot \roh_b(0)$. We are interested in the reduced density matrix $\roh_a (t)=tr_b(\roh(t))$ with components $\bra{n}\roh_a(t)\ket{m}$. Let us assume the oscillators are initially prepared in thermal states
\be\label{Th1}
\roh(0)=\frac{1}{z_a}\,e^{-\beta_1\hbar\om_0\,\ahd\ah}\ot \frac{1}{z_b}\,e^{-\beta_2\hbar\om_0\,\bhd\bh},
\ee
where $z_{a(b)}$ and $\beta_{1(2)}$ are the partition functions and inverse temperatures of the oscillators $a$ and $b$, respectively. With this choice, we find that the reduced density matrix $\roh_a (t)$ is a diagonal matrix with the diagonal elements
(see Appendix)
\be\label{R2}
P^a_{n} (t)=\bra{n}\roh_a(t)\ket{n}=\frac{[|f(t)|^2\,\bar{n}_a+|h(t)|^2\,\bar{n}_b]^n}{[|f(t)|^2\,\bar{n}_a+|h(t)|^2\,\bar{n}_b+1]^{n+1}},
\ee
where we defined
\beq\label{def1}
&&\bar{n}_{a(b)}=\frac{1}{e^{\beta_{1(2)}\hbar\omega_0}-1},\nn\\
&&G(t)=\int_0^t dt'\,g(t'),\nn\\
&&f(t)=e^{-i\omega_0 t}\cos(G(t)),\nn\\
&&h(t)=-ie^{-i\omega_0 t}\sin(G(t)).
\eeq
By making use of the Bogoliubov transformations (\ref{Bogo}) and Hamiltonian (\ref{HAB}), one finds (see Appendix)
\beq\label{Heisaad}
\hat{a} (t)=&& f(t)\hat{a}+ h(t)\hat{b},\nn\\
\hat{a}^\dag (t)=&& \bar{f}(t) \hat{a}^\dag +\bar{h} (t) \hat{b}^\dag.
\eeq
In the long-time regime $(t\gg \tau)$, thermalization occurs and (\ref{R2}) should tend to the following limit
\be\label{Th2}
P^a_n (\infty)=\frac{\bar{n}_b^n}{(\bar{n}_b+1)^{n+1}},
\ee
therefore,
\begin{eqnarray}\label{cossin}
&& \lim_{t\rightarrow\infty}|f(t)|^2=\lim_{t\rightarrow\infty}\cos^2(G(t))=0,\nn\\
&& \Rightarrow \lim_{t\rightarrow\infty}|h(t)|^2=\lim_{t\rightarrow\infty}\sin^2(G(t))=1.
\end{eqnarray}
For a cooling process, let's assume the bath oscillator ($b$-oscillator) is held at zero temperature. In this case $\bar{n}_b=0$, and we find from (\ref{R2})
\be
P^a_n (t)=\frac{(\cos^2 (G(t))\,\bar{n}_a)^n}{(\cos^2(G(t))\,\bar{n}_a+1)^{n+1}}.
\ee
In the long-time regime, using (\ref{cossin}) we find $P^a_n (\infty)=\delta_{n,0}$, indicating that the $a$-oscillator will finally fall in its ground state, as expected. For a dissipative system, the coupling function is a monotonically decreasing function of time. Therefore, the function $G(t)$ is an increasing function, and from condition (\ref{cossin}), we deduce that $\lim\limits_{t\rightarrow\infty}G(t)=G(\infty)$ should be the first root of the cosine function. Therefore, for the choice $g(t)=g_0 e^{-\gamma t}$, we have
\beq\label{Gt}
\cos(G(t))=\cos\left[\frac{g_0}{\gamma}(1-e^{-\gamma t})\right]&=&\cos\left[\frac{\pi}{2}\,(1-e^{-\gamma t})\right],\nn\\
                                                               &=&\sin\left[\frac{\pi}{2} e^{-\gamma t}\right].
\eeq
In Fig.~(\ref{comparison}), the functions $g(t)/g(0)=e^{-\gamma t}$ and $\cos^{2} G(t)$ are depicted in terms of $\tau=\gamma t$, which shows that the behavior of these functions is decreasing and nearly identical.
For an exponential decay, with dissipation coefficient $\gamma$, if we set
\be\label{choice}
\cos^2 (G(t))=e^{-\gamma t},
\ee
then the results obtained here for a dissipative oscillator will agree with those obtained from the Lindblad master equation.

\begin{figure}[t]
    \centering
    \includegraphics[width=.4\textwidth]{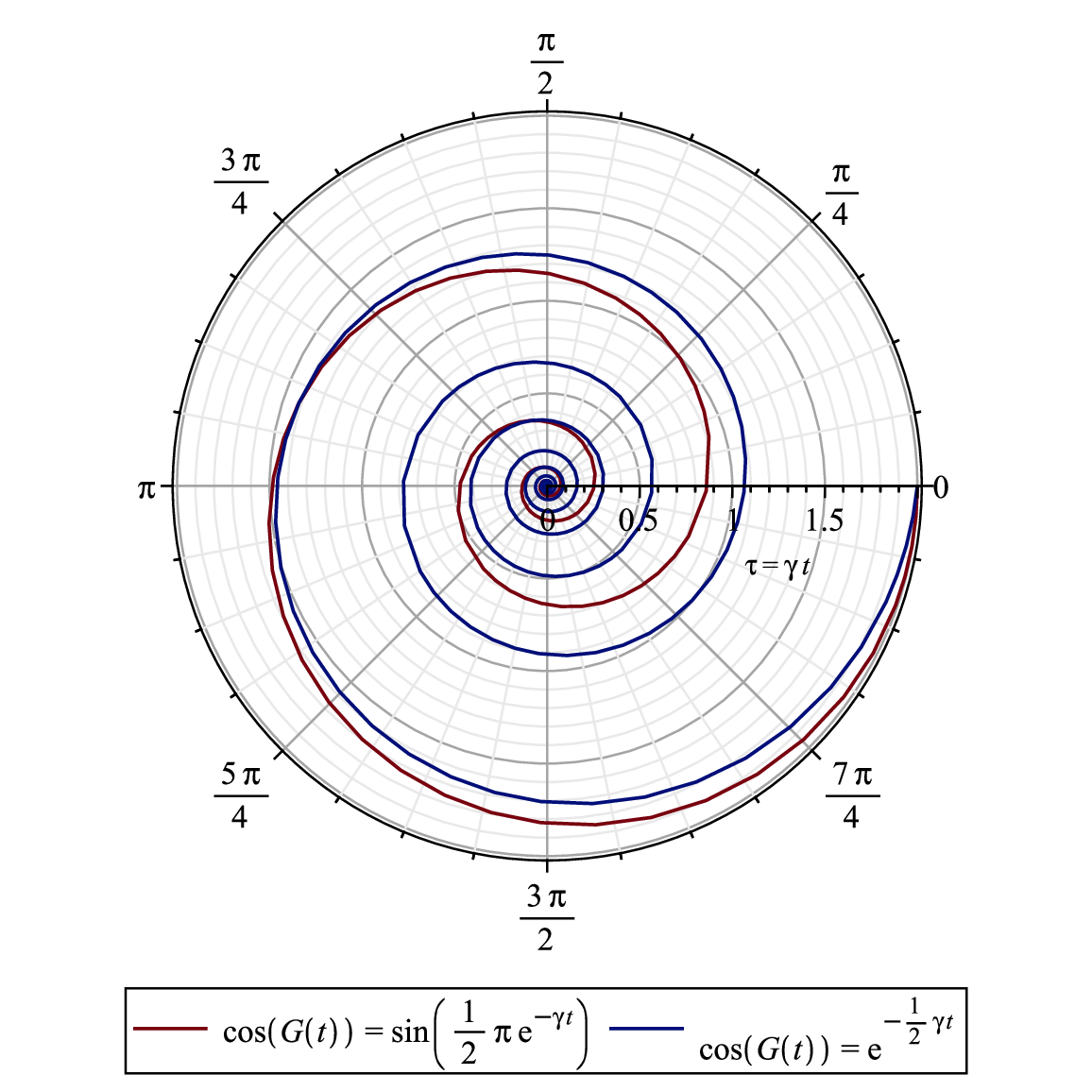}
    \caption{(Color online) The locus of the maximum of the Husimi distribution function on phase space in terms of the dimensionless parameter $\tau=\gamma t$ for the choices (\ref{Gt},\ref{choice}) and values $\alpha_0=2$ and $\gamma=0.2\,\omega_0$. }\label{curve}
\end{figure}

\begin{figure}[t]
    \centering
    \includegraphics[width=.5\textwidth]{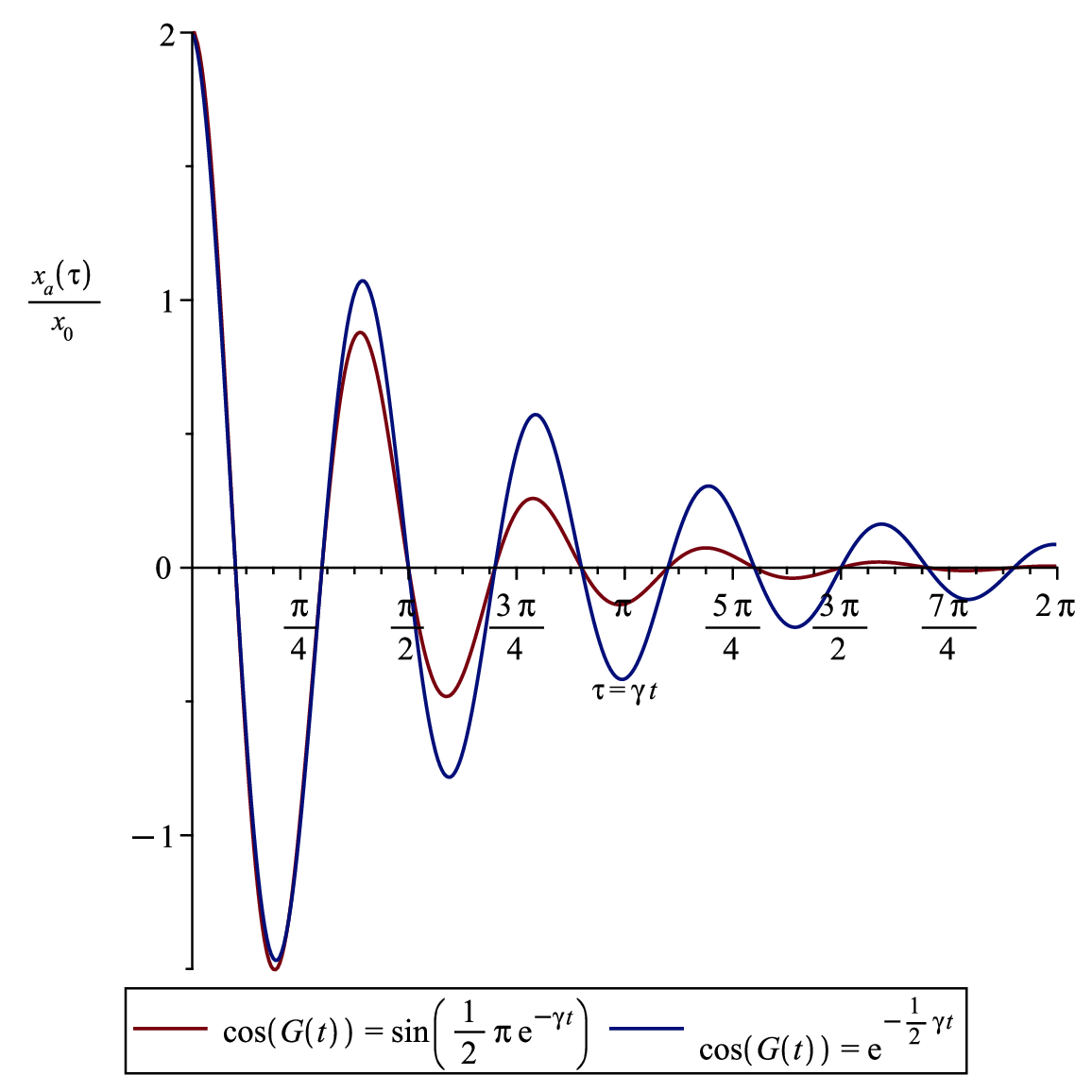}
    \caption{The dimensionless position in terms of the dimensionless parameter $\tau=\gamma t$ for the choices (\ref{Gt},\ref{choice}) and values $\alpha_0=2$ and $\gamma=0.2\,\omega_0$. }\label{osc}
\end{figure}

\begin{figure}[t]
    \centering
    \includegraphics[width=.5\textwidth]{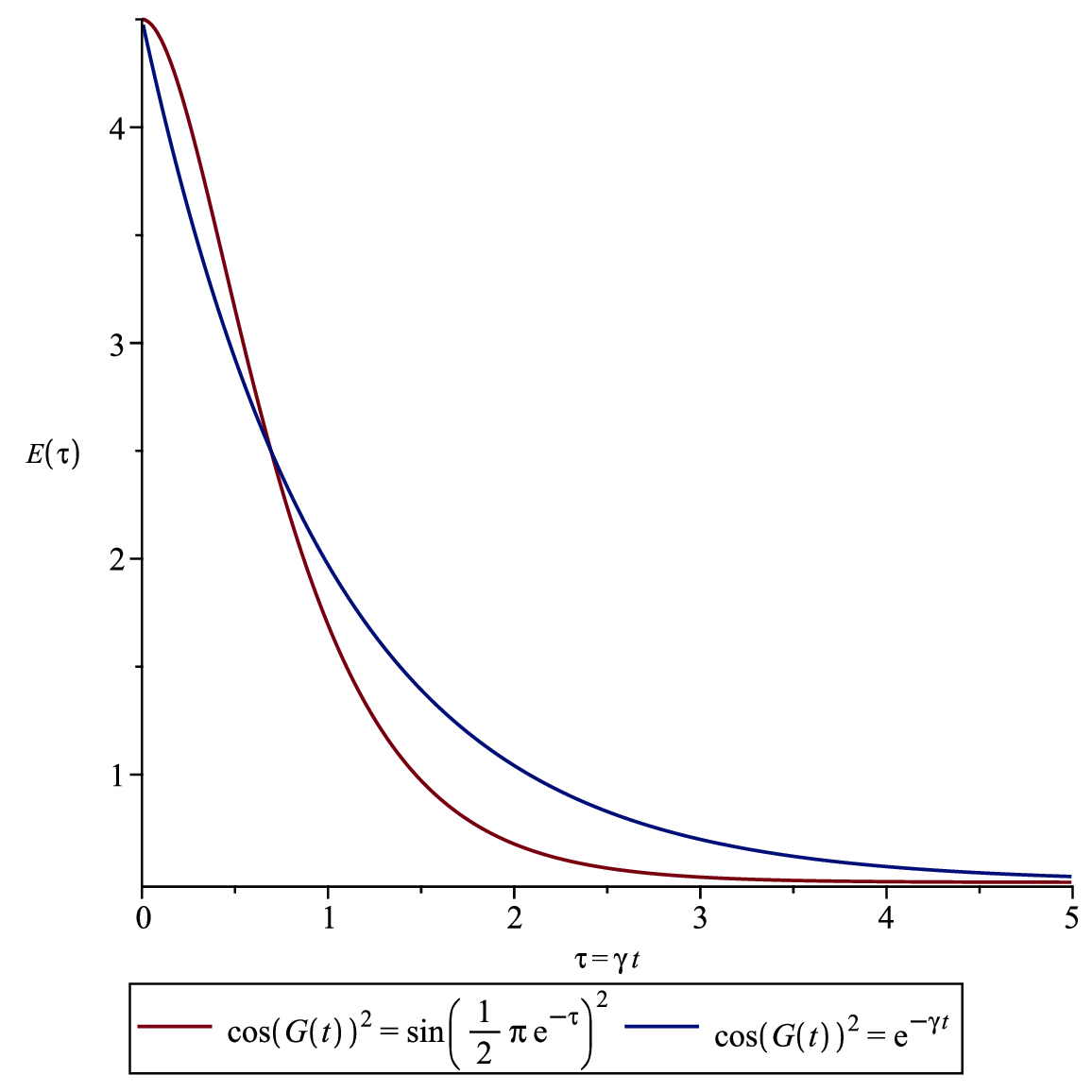}
    \caption{(Color online) The energy of the oscillator in terms of the dimensionless parameter $\tau=\gamma t$ for the choices (\ref{Gt},\ref{choice}), and $\alpha=2,\,\,\bar{n}_b=0$.}\label{Energy}
\end{figure}

Now suppose that the main oscillator is initially prepared in a coherent state and the bath-oscillator in a thermal state, then
\begin{equation}\label{coherentthermal}
\hat{\rho} (0)=\ket{\alpha_0}\bra{\alpha_0}\otimes\frac{e^{-\beta\hbar\omega_0\,\hat{b}^\dag \hat{b}}}{z_b}.
\end{equation}
By tracing out the bath oscillator degrees of freedom, we find the Husimi distribution function \cite{Schleich2011} corresponding to the reduced density matrix of the main oscillator at time
$t$ as (see Appendix)
\begin{eqnarray}\label{Husimi}
Q(\alpha,t)&=& \frac{1}{\pi}\bra{\alpha}\hat{\rho}_a (t)\ket{\alpha},\nn\\
         &=& \frac{1}{\pi}\frac{1}{1+|h(t)|^2\,\bar{n}_b}\,e^{-\frac{|\alpha-f(t)\alpha_0|^2}{1+|h(t)|^2\,\bar{n}_b}}.
\end{eqnarray}
Therefore, the maximum of $Q(\alpha,t)$ moves on the path $\alpha(t)=\alpha_0\,e^{-i\omega_0 t}\cos(G(t))$ on the complex plane or the phase space, see Fig (\ref{curve}).
The position of the main oscillator can be obtained as
\begin{eqnarray}\label{position}
x_a(t) &=& tr(\rho(0)\hat{x}(t))=\sqrt{\frac{\hbar}{2m\omega}}tr(\rho(0)[\hat{a}^\dag (t)+\hat{a}(t)]),\nonumber\\
       &=& \sqrt{\frac{\hbar}{2m\omega_0}}\,(f(t)\alpha_0+\bar{f}(t)\bar{\alpha}_0),\nonumber\\
       &=& \sqrt{\frac{2\hbar}{m\omega_0}}|\alpha_0|\cos(G(t))\cos(\omega_0 t-\theta),
\end{eqnarray}
where we assumed $\alpha_0=|\alpha_0|e^{i\theta}$ and made use of (\ref{choice}, \ref{coherentthermal}), and the following relations
\begin{eqnarray}
tr_b (\frac{e^{-\beta\hbar\omega_0\hat{b}^\dag\hat{b}}}{z_b}\hat{b})=tr_b(\frac{e^{-\beta\hbar\omega_0\hat{b}^\dag\hat{b}}}{z_b}\hat{b}^\dag)=0.
\end{eqnarray}
In Fig.~(\ref{osc}), the dimensionless position $x_a(t)/x_0$, ($x_0=\sqrt{2\hbar/m\omega_0}$) is depicted for $\alpha_0=2$, and $\gamma=0.2\omega_0$, which is compatible with the results depicted in Fig.~(\ref{curve}).
The energy of the main oscillator can also be calculated as
\beq\label{energy}
E(t)&=& tr[\hbar\omega_0(\hat{a}^\dag(t)\hat{a}(t)+1/2)\rho(0)],\nn\\
    &=& \frac{\hbar\omega_0}{2}+\hbar\omega_0 (|f(t)|^2|\alpha|^2+|h(t)|^2\bar{n}_b),\nn\\
    &=& \frac{\hbar\omega_0}{2}+\hbar\omega_0 \big(|\alpha|^2\cos^2(G(t))+\bar{n}_b \sin^2(G(t))\big),
\eeq
which decays exponentially to the expected value $E(\infty)=\frac{\hbar\omega_0}{2}\coth(\frac{\beta\hbar\omega_0}{2})$ for $G(t)$ given by (\ref{Gt}). In Fig.~(\ref{Energy}), the energy of the oscillator is depicted for choices (\ref{Gt}, \ref{choice}), and values $\alpha=2,\,\,\bar{n}_b=0$.
\subsubsection{Heat distribution of the $a$-oscillator}
\noindent In the realm of quantum thermodynamics \cite{Defner2019}, we revisit the Hamiltonian (\ref{H1}) to explore the heat distribution and its characteristic function for the $a$-oscillator, which is coupled to a heat bath with inverse temperature $\beta_2$, while itself maintained at an inverse temperature $\beta_1$. The heat distribution function is defined as follows \cite{Talkner2007,Esposito2009,Campisi2011,Whitney2014,Brander2015,Denzler2018,Levy2020,Funo2018}
\be\label{dis1}
P (Q,t)=\sum_{n=0}^\infty\sum_{m=0}^\infty \delta(Q-[E_m-E_n])\,P^a_{n\rightarrow m} (t)\,P^a_n (0),
\ee
where $E_n$'s are the energy levels of the $a$-oscillator, $P^a_n (0)=e^{-\beta_1\hbar\om_0\,n}/z_a$ denotes the probability of the $a$-oscillator being initially at the state $\ket{n}$, and $P^a_{n\rightarrow m} (t)=|\bra{m}\Uh (T)\ket{n}|^2$ signifies the transition probability at time $t$ for the process $\ket{n}\rightarrow\ket{m}$. To avoid the complexities associated with the Dirac delta function, it is simpler to use the characteristic function
\be\label{chr1}
G(\mu,t)=\int dQ\,e^{i\mu\,Q}\,P(Q,t).
\ee
After straightforward calculations, we find (see Appendix)
$$ G(\mu,t)=\frac{(e^{\beta_1\hbar\om_0}-1)\,e^{i\mu\hbar\om_0}}{e^{i\mu\hbar\om_0}[(1+|h|^2\bar{n}_b)e^{\beta_1\hbar\om_0}-|h|^2\bar{n}_be^{(\beta_1+i\mu)\hbar\om_0}+|h|^2(\bar{n}_b+1)-1]-|h|^2(\bar{n}_b+1)}.$$
\beq\label{G1}
\eeq
The characteristic function (\ref{G1}) coincides with the result reported in \cite{Denzler2018} obtained from Lindblad master equation \cite{Lindblad1976} for the choice (\ref{choice}).
\section{An oscillator interacting with two baths at different temperatures}
\begin{figure}[t]
    \centering
    \includegraphics[width=.4\textwidth]{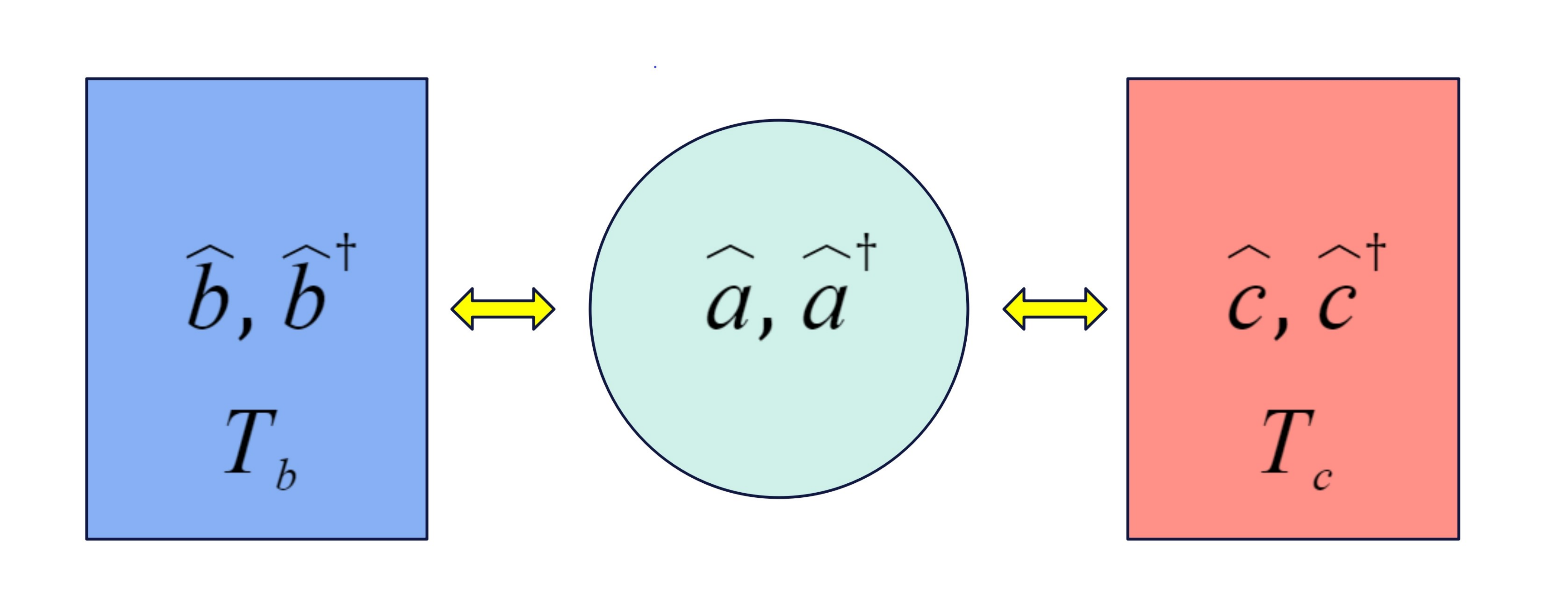}
    \caption{(Color online) An oscillator interacting with two baths at different temperatures.}\label{figure2}
\end{figure}
\noindent The extension of the scheme to multiple baths is straightforward. For instance, let's consider an oscillator interacting with two independent baths at inverse temperatures $\beta_1$ and $\beta_2$, as shown in Fig.~(\ref{figure2}). In this scenario, the total Hamiltonian is given by
\be\label{two1}
\Hh (t)=\hbar\om\,(\ahd\ah+\bhd\bh+\chd\ch)+\hbar g(t)\,(\ah\bhd+\ahd\bh+\ah\chd+\ahd\ch).
\ee
Solving the Heisenberg equations of motion yields (see Appendix)
\beq\label{two2}
\ah (t)=&& e^{-i\om t}\,\cos(\sqrt{2}\,G(t))\,\ah (0)\nn\\
&&+\frac{i\sqrt{2}}{2}\,e^{-i\om t}\,\sin(\sqrt{2}\,G(t))\,(\bh (0)+\ch (0)).\nn\\
\eeq
Suppose the $a$-oscillator is initially in an arbitrary state $\rho_a (0)$; then, the energy of the $a$-oscillator at time $t$ is given by
\beq\label{two3}
\ave{\Hh_a}_t &=& tr(\roh (0)\hbar\om\ahd (t)\ah(t)),\nn\\
&=&\hbar\om\cos^2 (\sqrt{2} G(t))\ave{\ahd (0)\ah (0)}\nn\\
&&\,\,\,+\frac{\hbar\om}{4}\,\sin^2 (\sqrt{2} G(t))\,\Big(\coth(\beta_1\hbar\om/2)+\coth(\beta_2\hbar\om/2)\Big).\nn\\
\eeq
Now, by selecting the coupling function as $g(t)=g_0 e^{-\gamma t}$ and applying the thermalization condition, we set $g_0/\gamma=\pi/2\sqrt{2}$. Therefore, $\cos^2 (\sqrt{2} G(t))=\sin^2(\frac{\pi}{2}e^{-\gamma t})$, $\sin^2 (\sqrt{2} G(t))=\cos^2(\frac{\pi}{2}e^{-\gamma t})$, and (\ref{two3}) can be rewritten as
\beq\label{two4}
\ave{\Hh_a}_t &=& tr(\roh (0)\hbar\om\ahd (t)\ah(t)),\nn\\
&=&\hbar\om\sin^2\left(\frac{\pi}{2}e^{-\gamma t}\right)\ave{\ahd (0)\ah (0)}\nn\\
&&\,\,\,+\frac{\hbar\om}{4}\,\cos^2\left(\frac{\pi}{2}e^{-\gamma t}\right)\,\Big(\coth(\beta_1\hbar\om/2)+\coth(\beta_2\hbar\om/2)\Big).\nn\\
\eeq
In the long-time regime ($\gamma t\gg 1$), we have
\be\label{two4}
\ave{\Hh_a}_t=\frac{\hbar\om}{4}\,\Big(\coth(\beta_1\hbar\om/2)+\coth(\beta_2\hbar\om/2)\Big).
\ee
In the classical regime ($\beta_{1(2)}\hbar\om\ll 1$), we obtain
\be\label{two5}
\ave{\Hh_a}_t=\kappa_B\,\Big(\frac{T_1+T_2}{2}\Big),
\ee
as expected.
\section{Dissipative two-level system}
\begin{figure}[t]
    \centering
    \includegraphics[width=.4\textwidth]{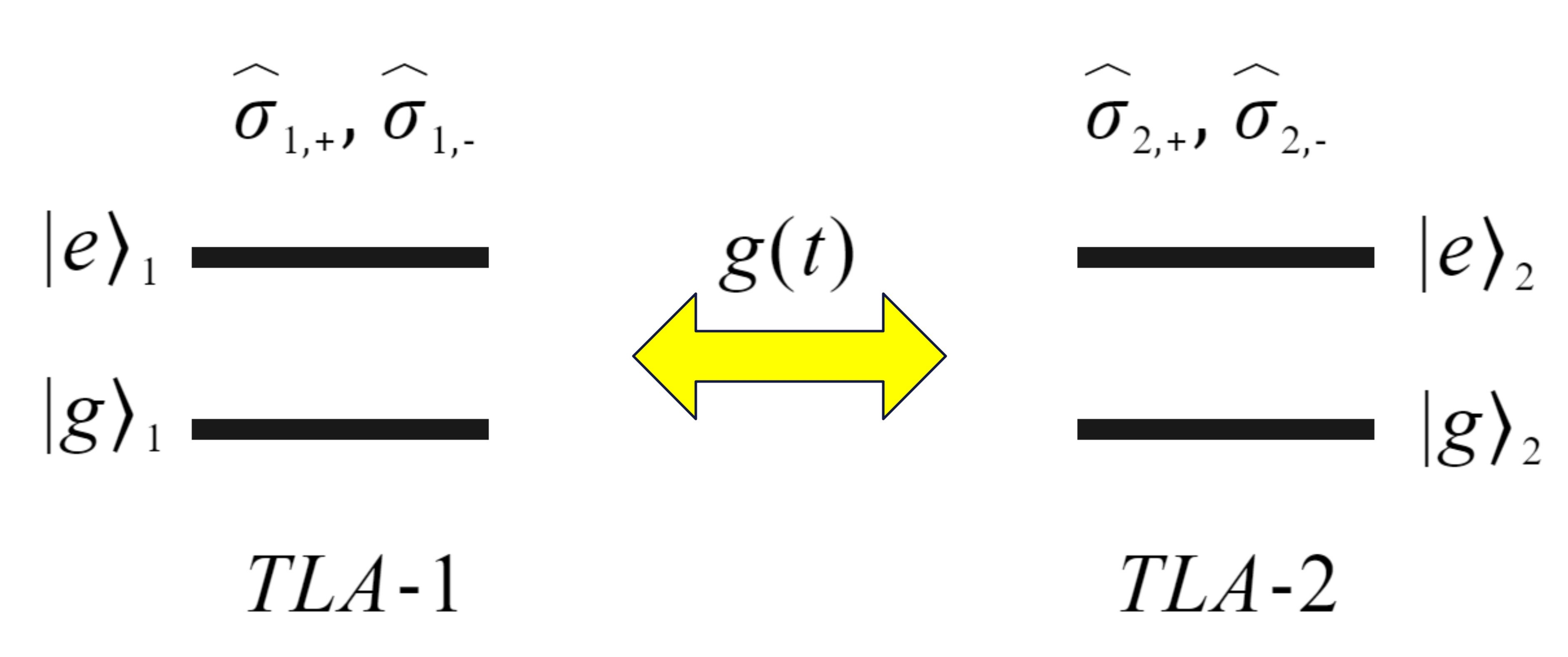}
    \caption{(Color online) A two-level atom in a dissipative environment. }\label{figure4}
\end{figure}
\noindent For a dissipative two-level system, we opt to model the bath system as a replica of the main system, as depicted in Fig.~(\ref{figure4}). Alternatively, we could couple this system to a single bosonic mode, which we will explore in the next section when discussing the pure dephasing model. In both scenarios, the systems interact linearly through a time-dependent coupling function $g(t)$, and the total Hamiltonian is described by
\be\label{TL1}
\Hh (t)=\frac{\hbar\om_0}{2}\,(\sigma^1_z+\sigma^2_z)+\hbar\,g(t)\,(\sigma^1_{-}\sigma^2_{+}+\sigma^1_{+}\sigma^2_{-}).
\ee
Let us consider the state of the total system at time $t$ in the standard basis as
\beq\label{TL2}
\ket{\psi(t)}=&&C_{++}\ket{+}_1\ket{+}_2+C_{+-}\ket{+}_1\ket{-}_2\nn\\
&&+C_{-+}\ket{-}_1\ket{+}_2+C_{--}\ket{-}_1\ket{-}_2,
\eeq
where $C_{\pm\pm}$ are the probability amplitudes. By solving the Schr\"{o}dinger equation we find the evolution matrix in the standard basis as (see Appendix)
\be\label{TL3}
\hat{W} (t)=\left(
          \begin{array}{cccc}
            e^{-i\om_0 t} & 0 & 0 & 0 \\
            0 & \cos(G(t)) & -i\sin(G(t)) & 0 \\
            0 & -i\sin(G(t)) & \cos(G(t)) & 0 \\
            0 & 0 & 0 & e^{i\om_0 t} \\
          \end{array}
        \right).
\ee
Now, let us assume that the bath is initially prepared in the thermal state
\be\label{TL5}
\roh_2 (0)=\left(
             \begin{array}{cc}
               p_e & 0 \\
               0 & p_g \\
             \end{array}
           \right), \,\,\,(p_e+p_g=1),
\ee
and the main system is initially prepared in an arbitrary state
\be\label{TL4}
\roh_1 (0)=\left(
             \begin{array}{cc}
               a & \bar{c} \\
               c & b \\
             \end{array}
           \right),\,\,\,\,(a+b=1,\,\,(a-b)^2+4c\bar{c}\leq 1),
\ee
therefore,
\beq\label{TL6}
\roh_1 (0)\ot\roh_2 (0)=\left(
                          \begin{array}{cccc}
                            a p_e & 0 & \bar{c} p_e & 0 \\
                            0 & a p_g & 0 & \bar{c} p_g \\
                            c p_e & 0 & b p_e & 0 \\
                            0 & c P_g & 0 & b p_g \\
                          \end{array}
                        \right).\nn\\
\eeq
The total density matrix at time $t$ is $\roh (t)=\Uh (t)\roh (0)\Uhd (t)$, and by taking the partial trace over the bath degrees of freedom, we find (see Appendix)
\beq\label{TL7}
\roh_1 (t)=\left(
             \begin{array}{cc}
               \phi(t) & \bar{c}\,e^{-i\om_0 t}\,\cos(G(t)) \\
               c\,e^{i\om_0 t}\,\cos(G(t)) & \psi(t) \\
             \end{array}
           \right),
\eeq
where for notational simplicity, we defined $\phi(t)=a\cos^2(G(t))+p_e\,\sin^2(G(t))$ and $\psi(t)=b\cos^2(G(t))+p_g\,\sin^2(G(t))$. In the long-time regime the main system is thermalized
\be\label{TL8}
\roh_1 (\infty)=\left(
             \begin{array}{cc}
               p_e & 0 \\
               0 & p_g \\
             \end{array}
           \right),
\ee
and when the bath is held in its ground state (\(\roh_2 (0)=\ket{g}\bra{g}\)), we have
\be\label{TL9}
\roh_1 (t)=\left(
             \begin{array}{cc}
               a\,\cos^2 (G(t)) & \bar{c}\,e^{-i\om_0 t}\,\cos(G(t)) \\
               c\,e^{i\om_0 t}\,\cos(G(t)) & b+a\,\sin^2 (G(t)) \\
             \end{array}
           \right),
\ee
therefore, the population decays as $a\,\cos^2 (G(t))$. For the particular choice (\ref{choice}), we have $\roh_{1,ee} (t)=a\,e^{-\gamma t}$ (see Fig.~\ref{comparison}), and $\roh_{1,gg} (t)=1-a\,e^{-\gamma t}$, indicating the spontaneous emission for $a=1$ and $b=0$, \cite{Scully1996}.
\subsection{Pure dephasing model}
\begin{figure}[t]
    \centering
    \includegraphics[width=.5\textwidth]{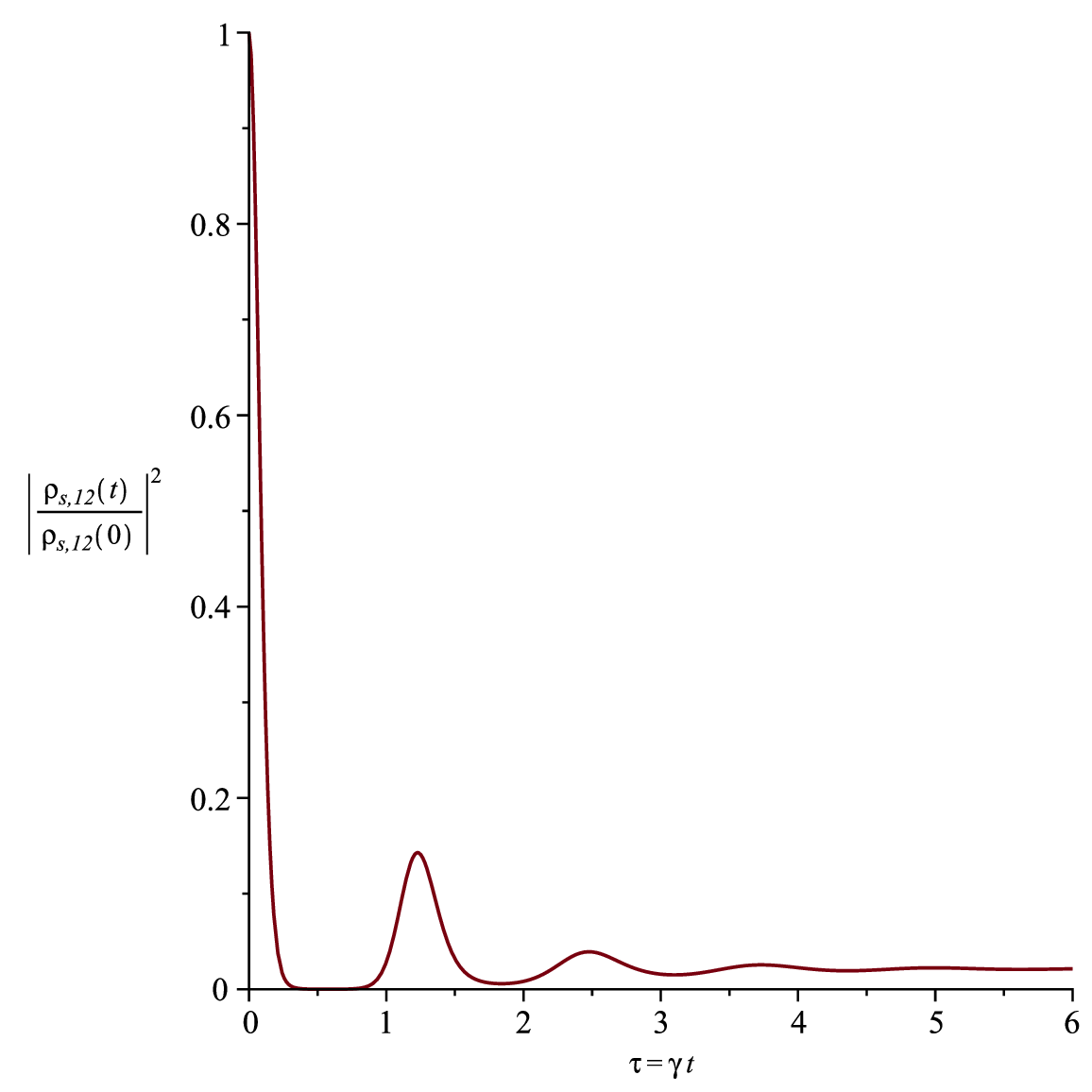}
    \caption{(Color online) The scaled norm of the off-diagonal element $|\hat{\rho}_{s,12}/\hat{\rho}_{s,12}(0)|^2$ in terms of $\tau=\gamma t$ for the choice $g(t)=g_0\,e^{-\gamma t}$ and values $\gamma=0.2\,\omega,\,g_0=\omega,\,\,\beta=\infty$.}\label{figure5}
\end{figure}

\noindent For pure dephasing model \cite{Scully1996,Walls2008,Manfredi2023}, let us couple the main two-level system to a single bosonic mode as follows
\begin{equation}\label{dephase1}
  \hat{H} (t)=\frac{\hbar\omega_0}{2}\,\sigma_z+\hbar\omega\,\hat{b}^\dag \hat{b}+\hbar g(t)\,\sigma_z\otimes (\hat{b}+\hat{b}^\dag).
\end{equation}
In the interaction picture, we have
\begin{equation}\label{dephase2}
  \hat{H}_I (t)=\hbar g(t)\,\sigma_z\otimes (e^{-i\omega t}\hat{b}+e^{i\omega t}\hat{b}^\dag).
\end{equation}
Since the Hamiltonian $\hat{H}_I (t)$ commutes at different times $[\hat{H}_I (t),\hat{H}_I (t')]=0$, the evolution operator in the interaction picture can be obtained in closed form as (see Appendix)
\begin{equation}\label{dephase3}
  \hat{U}_I (t)=i_d\otimes\cosh(\hat{A})+\sigma_z\otimes\sinh(\hat{A}),
\end{equation}
where
\begin{eqnarray}\label{dephase4}
&& \hat{A}=\xi(t)\hat{b}-\bar{\xi}(t)\hat{b}^\dag,\\
&& \xi(t)=-i\int_0^t dt'\,g(t')\,e^{-i\omega t'},
\end{eqnarray}
and $i_d$ is the identity operator on the Hilbert space of the two-level system. The evolution operator in Schr\"{o}dinger picture is $\hat{U}(t)=\hat{U}_0 (t)\hat{U}_I (t)$ where
\begin{equation}\label{uzero}
\hat{U}_0 (t)=e^{-\frac{i\omega_0 t}{2}\sigma_z}\otimes e^{-i\omega t\,\hat{b}^\dag\hat{b}}.
\end{equation}
Let the two-level system be initially prepared in an arbitrary state and the bath be hold in a thermal state with inverse temperature $\beta$
\begin{equation}\label{dephase4}
  \hat{\rho}(0)=\underbrace{\left(
                  \begin{array}{cc}
                    a & \bar{c} \\
                    c & b \\
                  \end{array}
                \right)}_{ \hat{\rho}_s (0)}\otimes\frac{e^{-\beta\hbar\omega\,\hat{b}^\dag \hat{b}}}{z_b}.
\end{equation}
After tracing out the bath degrees of freedom, the reduced density matrix of the main system at an arbitrary time $t$ can be obtained as (see Appendix)
\begin{eqnarray}\label{dephase5}
  \hat{\rho}_s (t)=\left(
                     \begin{array}{cc}
                       a & e^{-2 |\xi(t)|^2\coth(\beta\hbar\omega/2)}e^{-i\omega_0 t}\,\bar{c} \\
                       e^{-2 |\xi(t)|^2\coth(\beta\hbar\omega/2)}e^{i\omega_0 t}\,c & b \\
                     \end{array}
                   \right).\nn\\
\end{eqnarray}
Therefore, the diagonal elements remain unchanged as expected, while the off-diagonal elements decrease by increasing temperature. For the choice $g(t)=g_0\,e^{-\gamma t}$, the scaled norm of the off-diagonal element
\beq\label{offdiag}
\left|\frac{\hat{\rho}_{s,12}(t)}{\hat{\rho}_{s,12}(0)}\right|^2 &=& e^{-4\coth(\frac{\beta\hbar\omega)}{2}|\xi(t)|^2},\nn\\
                                           &=& e^{\frac{-4g_0^2}{\gamma^2+\omega^2}\left[1+e^{-2\gamma t}-2e^{-\gamma t}\cos(\omega t)\right]\coth(\beta\hbar\omega/2)},
\eeq
is depicted in terms of the dimensionless variable $\tau=\gamma t$ for values $g_0=\omega,\,\gamma=0.2\omega,\,\,\beta=\infty$, in Fig.~(\ref{figure5}). In the long-time regime, and high temperature limit ($\beta\rightarrow 0$), the coherency decays in temperature as $e^{-\frac{8g_0^2\,\kappa_B}{(\gamma^2+\omega^2)\hbar\omega}T}$.
%
\subsection{Markovian and non-Markovian process}
\noindent A measure for the distance between two density matrices $\roh_1$ and $\roh_2$ is defined by \cite{Nielsen2010}
\be\label{distance}
D(\roh_1,\roh_2)=\frac{1}{2}\, tr(|\roh_1-\roh_2|),
\ee
where $|\hat{A}|=\sqrt{\hat{A}^{\dag}\hat{A}}$. For a Hermitian matrix ($\hat{A}^\dag=\hat{A}$), the trace $tr(|\hat{A}|)$ can be written as a sum over the absolute values of the eigenvalues of $\hat{A}$, $tr(|\hat{A}|)=\sum\limits_i |\lambda_i|$. The rate of change of the distance of evolved density matrices is defined by
\be\label{rated}
\sigma(t,\roh_1(0),\roh_2(0))=\frac{d}{dt}D(\roh_1 (t),\roh_2 (t)).
\ee
A process is said to be non-Markovian \cite{Gardiner2004,Rivas2014,Laine2014,Budini2014,Breuer2016,Vega2017,Tamascelli2018} if there exists a pair of initial states $\roh_1 (0)$ and $\roh_2 (0)$ and a certain time $t$ such that $\sigma(t,\roh_1(0),\roh_2(0))>0$, \cite{Laine2014}. In physical terms, this implies that during non-Markovian dynamics, the ability to distinguish between the two states improves at specific moments. This can be understood as information flowing from the environment back into the system, which increases the likelihood of differentiating the two states.

For the dissipative two-level atom let us choose the arbitrary initial states as
\be\label{rozero}
\roh_1 (0)=\left(
             \begin{array}{cc}
               a_1 & \bar{c}_1 \\
               c_1 & b_1 \\
             \end{array}
           \right),\,\,\,\roh_2 (0)=\left(
                                      \begin{array}{cc}
                                        a_2 & \bar{c}_2 \\
                                        c_2 & b_2 \\
                                      \end{array}
                                    \right),
\ee
where consistency conditions given by (\ref{TL4}) are assumed. By making use of (\ref{TL7}), the distance between the evolved states can be obtained as
\be\label{TDTLS}
 D(\roh_1 (t),\roh_2 (t))=\sqrt{|a_1-a_2|^2\cos^4 (G(t))+|c_1-c_2|^2\cos^2 (G(t))}.
\ee
The rate of the trace distance (\ref{TDTLS}) is
\beq\label{DRate}
&& \sigma(t,\roh_1(0),\roh_2(0))= \frac{d}{dt}D(\roh_1 (t),\roh_2 (t)),\nn\\
&&                             = -\frac{g(t)\sin[2G(t)]\big[2|a_1-a_2|^2\cos^2(G(t))+|c_1-c_2|^2\big]}{\big[|a_1-a_2|^2\cos^4 (G(t))+|c_1-c_2|^2\cos^2 (G(t))\big]^{3/2}}.
\eeq
Therefore, the sign of $\sigma(t,\roh_1(0),\roh_2(0))$ is determined by the term $g(t)\sin[2G(t)]$. For the choices (\ref{Gt},\ref{choice}), we have $g(t)\sin[2G(t)]\geq 0$, so in any time interval we have $\sigma(t,\roh_1(0),\roh_2(0))\leq 0$, and the process is Markovian.
\section{Conclusion}
\noindent In this article, we have introduced a novel scheme for investigating dissipation or thermalization in open quantum systems. This method is based on time-dependent coupling functions, and by appropriately choosing these functions, the effects of dissipation can be modeled on the primary quantum system. To demonstrate the efficiency and simplicity of the method, we examined the quantum dynamics of a harmonic oscillator interacting with a thermal bath for different initial conditions and obtained accurate results for reduced density matrix elements (\ref{R2}), the Husimi distribution function (\ref{Husimi}), and its phase space representation (Fig.~\ref{curve}), as well as the characteristic heat distribution function for the oscillator (\ref{G1}). These results were consistent with those obtained from other methods such as the Lindblad master equation. To demonstrate the method's ability to generalize to multiple thermal baths, we investigated the interaction of the oscillator with two thermal baths at different temperatures and obtained compatible results. Furthermore, we examined a two-level atom with energy or phase dissipation and obtained precise results for the reduced density matrices in both cases (\ref{TL7}, \ref{dephase5}, Fig.~\ref{figure5}), demonstrating spontaneous emission and pure dephasing processes, respectively. Finally, for a two-level system, by considering the time-dependent strictly decreasing coupling function $g(t)=g_0\,e^{-\gamma t}$, we showed that the process is Markovian. This method can be applied to any desired quantum system in the presence of dissipation or studying thermalization, and the form of the time-dependent coupling functions can be determined based on experimental data. Due to the simplicity of the total Hamiltonian in this scheme, numerical methods will be very effective in cases where the problem is not integrable.

\newpage
\subsection{Appendices}
\subsection{Derivation of (\ref{R2}), (\ref{Heisaad})}
\noindent By using $\hat{\rho}_a (t)=tr_b(\hat{\rho} (t))$, we obtain
\beq\label{s1}
_a\bra{n}\hat{\rho}_a (t)\ket{m}_a &=& _a\bra{0}\frac{(\hat{a})^n}{\sqrt{n!}}\,\hat{\rho}_a (t)\,\frac{(\hat{a}^\dag)^m}{\sqrt{m!}}\ket{0}_a,\nn\\
                              &=&_a\bra{0}\,tr_b\big(\frac{1}{\sqrt{n! m!}}\,(\hat{a})^n \hat{\rho} (t) (\hat{a}^\dag)^m\big)\ket{0}_a,\nn\\
                              &=& tr_a \Big(\ket{0}_{aa}\bra{0}\,tr_b \Big[\frac{1}{\sqrt{n! m!}}\,(\hat{a})^n \hat{\rho} (t) (\hat{a}^\dag)^m\Big]\Big),
\eeq
by inserting the identity
\be\label{s2}
\ket{0}_{aa}\bra{0}=\sum_{s=0}^\infty \frac{(-1)^s}{s!}\,(\hat{a}^\dag)^s (\hat{a})^s,
\ee
into equation (\ref{s1}), we obtain
\beq\label{s3}
_a\bra{n}\hat{\rho}_a (t)\ket{m}_a &=& tr\Big[\frac{1}{\sqrt{n! m!}}\sum_{s=0}^\infty \frac{(-1)^s}{s!}\,(\hat{a}^\dag)^s (\hat{a})^s\,(\hat{a})^n \hat{\rho} (t) (\hat{a}^\dag)^m\Big],\nn\\
                                  &=& \frac{1}{\sqrt{n! m!}}\sum_{s=0}^\infty \frac{(-1)^s}{s!}\,tr\Big((\hat{a}^\dag)^{s+m} (\hat{a})^{s+n}\,\hat{\rho} (t)\Big).
\eeq
Now, using $\hat{\rho} (t)=\hat{U} (t)\hat{\rho} (0) \hat{U}^\dag (t)$ and the Heisenberg representations
\beq\label{s4}
\hat{a} (t)=\hat{U}^\dag (t) \hat{a} \hat{U} (t),\nn\\
\hat{a}^\dag (t)=\hat{U}^\dag (t) \hat{a}^\dag \hat{U} (t),
\eeq
we rewrite (\ref{s3}) as
\be\label{s5}
_a\bra{n}\hat{\rho}_a (t)\ket{m}_a=\frac{1}{\sqrt{n! m!}}\sum_{s=0}^\infty \frac{(-1)^s}{s!}\,tr\Big((\hat{a}^\dag (t))^{s+m} (\hat{a} (t))^{s+n}\,\hat{\rho} (0)\Big).
\ee
By making use of the Bogoliubov transformations (3), the Hamiltonian (4), and definitions $\Omega_{A(B)} (t)=\omega t\mp G(t)$ ($G(t)=\int_0^t dt'\,g(t')$), we obtain
\beq\label{s6}
 \hat{a} (t) &&= \frac{1}{\sqrt{2}}(\hat{A}(t)+\hat{B}(t))=\frac{1}{\sqrt{2}}(e^{-i\Omega_A (t)}\underbrace{\hat{A}(0)}_{\frac{\hat{a}-\hat{b}}{\sqrt{2}}}+e^{-i\Omega_B (t)}\underbrace{\hat{B}(0)}_{\frac{\hat{a}+\hat{b}}{\sqrt{2}}}),\nn\\
             &&= \underbrace{\frac{1}{2}\big(e^{-i\Omega_B (t)}+e^{-i\Omega_A (t)}\big)}_{f(t)=e^{-i\omega t}\cos(G(t))} \hat{a}+ \underbrace{\frac{1}{2}\big(e^{-i\Omega_B (t)}-e^{-i\Omega_A (t)}\big)}_{h(t)=-ie^{-i\omega t}\sin(G(t))}\hat{b},\nn
\eeq
therefore,
\beq
\hat{a} (t) =&& f(t)\hat{a}+h(t)\hat{b},\nn\\
\hat{a}^\dag (t)=&& \bar{f}(t) \hat{a}^\dag +\bar{h} (t) \hat{b}^\dag.\nn
\eeq
We now employ binomial expansions and assume $\hat{\rho} (0)=\hat{\rho}_a (0)\otimes\hat{\rho}_b (0)$. Consequently,
\beq\label{s7}
&& _a\bra{n}\hat{\rho}_a (t)\ket{m}_a = \nn\\
&& \frac{1}{\sqrt{n! m!}}\sum_{s=0}^\infty \frac{(-1)^s}{s!}\sum_{p=0}^{s+m}\sum_{q=0}^{s+n}{s+m\choose p}{s+n\choose q} (\bar{f}(t))^{s+m-p}\,(\bar{h} (t))^{p}\nn\\
    &&\times (f (t))^{s+n-q}\,(h(t))^q\, tr_a \Big[(\hat{a}^\dag)^{s+m-p}\,\hat{a}^{s+n-q}\,\hat{\rho}_a (0)\Big]\,tr_b \Big[(\hat{b}^\dag)^p \,\hat{b}^q\,\hat{\rho}_b (0)\Big].\nn\\
\eeq
For the scenario where both oscillators are initially prepared in thermal states with inverse temperatures $\beta_1$ and $\beta_2$, we have
\be\label{s8}
\hat{\rho} (0)=\frac{e^{-\beta_1\hbar\,\hat{a}^\dag \hat{a}}}{z_a}\otimes\frac{e^{-\beta_2\hbar\,\hat{b}^\dag \hat{b}}}{z_b},
\ee
now using
\beq\label{s9}
tr_b \Big[(\hat{b}^\dag)^p \,\hat{b}^q\,\hat{\rho}_b (0)\Big]&=& \frac{1}{z_b}\sum_{k=0}^{\infty}\, _b\bra{k} (\hat{b}^{\dag})^p (\hat{b})^q\ket{k}_b \,e^{-\beta_2\hbar\omega k},\nn\\
                                                             &=&  \frac{\delta_{pq}}{z_b}\,\sum_{k=0}^{\infty}\, _b\bra{k} (\hat{b}^\dag)^q (\hat{b})^q\ket{k}_b \,e^{-\beta_2\hbar\omega k},\nn\\
                                                             &=&  \frac{\delta_{pq}}{z_b}\,\sum_{k=q}^\infty \frac{k!}{(k-q)!}\,e^{-\beta_2\hbar\omega k},\nn\\
                                                             &=&  \frac{\delta_{pq}}{z_b}\,\sum_{k=0}^\infty \frac{(k+q)!}{k!}\,e^{-\beta_2\hbar\omega (k+q)},\nn\\
                                                             &=&  \delta_{pq}\,q!\,\bar{n}_b^q,\,\,\,\,\,(\bar{n}_b=\frac{1}{e^{\beta_2\hbar\omega}-1}),
\eeq
we find
\beq\label{s10}
_a\bra{n}\hat{\rho}_a (t)\ket{m}_a &=& \frac{\delta_{n m}}{n!}\sum_{s=0}^\infty \frac{(-1)^s}{s!}\sum_{q=0}^{s+n} {s+n\choose q}^2\,|f(t)|^{2(s+n-q)}\,|h(t)|^{2q}\nn\\
&&\times\,(s+n-q)!\,q!\,\bar{n}_a^{s+n-q}\,\bar{n}_b^q,\nn
\eeq
\[
 = \frac{\delta_{n m}}{n!}\sum_{s=0}^\infty \frac{(-1)^s}{s!}\sum_{q=0}^{s+n} \frac{[(s+n)!]^2}{q! (s+n-q)!}\,\Big(|f(t)|^2 \bar{n}_a\Big)^{s+n-q}\Big(|h(t)| \bar{n}_b\Big)^q,
\]
\[=  \frac{\delta_{n m}}{n!}\sum_{s=0}^\infty \frac{(-1)^s}{s!}\Big(|f(t)|^2 \bar{n}_a+|h(t)|^2 \bar{n}_b \Big)^{s+n},\]
\be
= \delta_{n m}\,\frac{[|f(t)|^2 \bar{n}_a+|h(t)|^2 \bar{n}_b]^n}{n!}\frac{n!}{[|f(t)|^2 \bar{n}_a+|h(t)|^2 \bar{n}_b+1]^{n+1}},
\ee
where we used
\be\label{s12}
\sum_{s=0}^\infty \frac{(-1)^s (s+n)!}{s!}(x)^{s}=\frac{n!}{(1+x)^{n+1}}.
\ee
Hence, the reduced density matrix is diagonal, with its diagonal elements determined by
\[
P^a_n (t)=\frac{[|f(t)|^2 \bar{n}_a+|h(t)|^2 \bar{n}_b]^n}{[|f(t)|^2 \bar{n}_a+|h(t)|^2 \bar{n}_b+1]^{n+1}}.
\]
\subsection{Derivation of (\ref{Husimi})}
\noindent The initial state is given by
\[
\hat{\rho} (0)=\ket{\alpha_0}_{aa}\bra{\alpha_0}\otimes \frac{e^{-\beta\hbar\omega\,\hat{b}^\dag \hat{b}}}{z_b}.
\]
The Husimi distribution function corresponding to the reduced density matrix is defined as
\beq\label{s15}
Q(\alpha,t)&=& \frac{1}{\pi}\,_a\bra{\alpha}\hat{\rho}_a (t)\ket{\alpha}_a,\nn\\
           &=& \frac{1}{\pi}\,_a\bra{0}\hat{D}^\dag_a (\alpha)\hat{\rho}_a (t)\hat{D}_a (\alpha)\ket{0}_a,\nn\\
           &=& \frac{1}{\pi}\, _a\bra{0}tr_b \big(\hat{D}^\dag_a (\alpha)\hat{\rho} (t)\hat{D}_a (\alpha)\big)\ket{0}_a,\nn\\
           &=& \frac{1}{\pi}\,tr_a \Big[\ket{o}_{aa}\bra{0}tr_b \big(\hat{D}^\dag_a (\alpha)\hat{\rho} (t)\hat{D}_a (\alpha)\big)\Big],\nn\\
           &=& \frac{1}{\pi}\,tr\Big[\ket{o}_{aa}\bra{0}\hat{D}^\dag_a (\alpha)\hat{\rho} (t)\hat{D}_a (\alpha)\Big],\nn\\
\eeq
where $\hat{D}_a (\alpha)=e^{\alpha\hat{a}^\dag-\bar{\alpha}\hat{a}}$ represents the displacement operator, which acts on the vacuum state to produce a coherent state for the $a$-oscillator, denoted by $\ket{\alpha}=\hat{D}_a (\alpha)\ket{0}_a$. By utilizing $\hat{\rho}(t)=\hat{U} (t)\hat{\rho}(0)\hat{U}(t)$ and equation (\ref{s2}), we obtain
\beq\label{s16}
Q(\alpha,t)&=& \frac{1}{\pi}\sum_{s=0}^\infty \frac{(-1)^s}{s!}\,tr\Big[(\hat{a}^\dag)^s (\hat{a})^s\,\hat{D}^\dag_a (\alpha)\hat{U} (t)\hat{\rho}(0)\hat{U}(t)\hat{D}_a (\alpha)\Big],\nn\\
           &=& \frac{1}{\pi}\sum_{s=0}^\infty \frac{(-1)^s}{s!}\,tr\Big[\hat{U}^\dag (t)\hat{D}_a (\alpha)(\hat{a}^\dag)^s (\hat{a})^s\hat{D}^\dag_a (\alpha)\hat{U} (t)\hat{\rho}(0)\Big],\nn\\
           &=& \frac{1}{\pi}\sum_{s=0}^\infty \frac{(-1)^s}{s!}\,tr\Big[(\hat{a}^\dag (t)-\bar{\alpha})^s\,(\hat{a} (t)-\alpha)^s\,\hat{\rho}(0)\Big].
\eeq
To proceed, we substitute equations (\ref{s6}) into equation (\ref{s16}), yielding
\beq\label{s17}
&& Q(\alpha,t)= \frac{1}{\pi}\sum_{s=0}^\infty \frac{(-1)^s}{s!}\,tr\Big[(\bar{f}\hat{a}^\dag+\bar{h}\hat{b}^\dag-\bar{\alpha})^s\,(f\hat{a}+h\hat{b}-\alpha)^s\,\hat{\rho}(0)\Big],\nn\\
&&           = \frac{1}{\pi}\sum_{s=0}^\infty \frac{(-1)^s}{s!}\sum_{p=0}^s\sum_{q=0}^s {s\choose p}{s\choose q}\bra{\alpha_0}(\bar{f}\hat{a}^\dag-\bar{\alpha})^{s-p}\,(f\hat{a}-\alpha)^{s-q}\ket{\alpha_0}\nn\\
&& \times\bar{h}^p h^q\,\underbrace{tr_b\Big((\hat{b}^\dag)^p \hat{b}^q\,\frac{e^{-\beta\hbar\omega\,\hat{b}^\dag \hat{b}}}{z_b}\Big)}_{\delta_{pq} q!\bar{n}_b^q},\nn\\
&&= \frac{1}{\pi}\sum_{s=0}^\infty \frac{(-1)^s}{s!}\sum_{p=0}^s {s\choose p}^2\,|f(t)\alpha_0-\alpha|^{2(s-p)}|h(t)|^{2p}\,p! (\bar{n}_b)^p,\nn\\
&&= \frac{1}{\pi}\sum_{s=0}^\infty (-1)^s \big(|h(t)|^2\,\bar{n}_b\big)^s\sum_{u=0}^s\frac{s!}{(u!)^2\,(s-u)!}\Big(\frac{|f(t)\alpha_0-\alpha|^2}{|h(t)|^2\,\bar{n}_b}\Big)^u,\nn\\
&&= \frac{1}{\pi}\sum_{s=0}^\infty (-1)^s \big(|h(t)|^2\,\bar{n}_b\big)^s\,L_s \Big(-\frac{|f(t)\alpha_0-\alpha|^2}{|h(t)|^2\,\bar{n}_b}\Big),\nn\\
\eeq
where we made use of the following series expansion for Laguerre polynomials $L_s(x)$
\be\label{s18}
L_s (x)=\sum_{u=0}^s\frac{s!}{(u!)^2\,(s-u)!}(-x)^u.
\ee
Finally, using the identity
\be\label{s19}
\sum_{s=0}^\infty (y)^s\,L_s (x)=\frac{e^{-\frac{xy}{1-y}}}{1-y},
\ee
we obtain
$$
Q(\alpha,t)=\frac{1}{\pi}\frac{1}{1+|h(t)|^2\,\bar{n}_b}\,e^{-\frac{|\alpha-f(t)\alpha_0|^2}{1+|h(t)|^2\,\bar{n}_b}}.
$$
\subsection{Derivation of (\ref{G1})}
\noindent For the $a$-oscillator We have
$$
P(Q,t)=\sum_{n=0}^\infty\sum_{m=0}^\infty \delta(Q-[E_m-E_n])\,P^a_{n \to m} (t)\,P^a_n (0),\nn\\
$$
where
$$
P^a_n (0)=\frac{e^{-\beta_1\hbar\omega n}}{z_A (\beta_1)},\nn
$$
and
\beq\label{s22}
P^a_{n\to m} (t)&=& |_a\bra{m}\hat{U}_a (t)\ket{n}_a|^2= _a\bra{m}\hat{U}_a (t) \underbrace{\ket{n}_{aa}\bra{n}}_{\hat{\rho}_a(0)}\hat{U}_a^\dag (t)\ket{m}_a,\nn\\
                &=& _a\bra{m}\hat{\rho}_a (t)\ket{m}_a,\nn\\
                &=& \frac{1}{m!}\sum_{s=0}^\infty \frac{(-1)^s}{s!}\sum_{q=0}^{s+m} {s+m\choose q}^2\,q!\,\bar{n}_b^q\,|f(t)|^{2(s+m-q)}|h(t)|^{2q}\nn\\
                &&\times\,_a\bra{n}(\hat{a}^\dag)^{s+m-q}(\hat{a})^{s+m-q}\ket{n}_a,\nn
\eeq
where we used equations (\ref{s7}) and (\ref{s9}). The characteristic function $G(\mu, t)$ is expressed as
$$
G(\mu,t)=\int dQ\,e^{i\mu Q}\,P(Q,t)=\sum_{n=0}^\infty \sum_{m=0}^\infty e^{i\mu (E_m-E_n)}\,P^a_{n \to m} (t)\,P^a_n (0),
$$
$$ =\sum_{m=0}^\infty \frac{e^{i\mu\hbar\omega m}}{m!}\sum_{s=0}^\infty \frac{(-1)^s}{s!}\sum_{q=0}^{s+m} {s+m\choose q}^2\,|f(t)|^{2(s+m-q)}\big(|h(t)|^2 \bar{n}_b\big)^q\,q!\,\bar{n}_b$$
\be\label{s23}
\,\,\,\,\,\,\,\,\times\underbrace{\sum_{n=0}^\infty e^{-i\mu\hbar\omega n}\frac{e^{-\beta_1\hbar\omega n}}{z_a (\beta_1)}\,_a\bra{n}(\hat{a}^\dag)^{s+m-q} (\hat{a})^{s+m-q}\ket{n}_a}_{I},
\ee
where $I$ in (\ref{s23}) is
\beq\label{s24}
I=&& \frac{1}{z_a (\beta_1)}\,\sum_{n=s+m-q}^\infty e^{-(\beta_1+i\mu)\hbar\omega n}\,\underbrace{_a\bra{n}(\hat{a}^\dag)^{s+m-q} (\hat{a})^{s+m-q}\ket{n}_a}_{\frac{n!}{(n-s-m+q)!}},\nn\\
 =&& \frac{1}{z_a (\beta_1)}\,\sum_{j=0}^\infty e^{-(\beta_1+i\mu)\hbar\omega (s+m-q+j)}\,\frac{(s+m-q+j)!}{j!},\nn\\
 =&& \frac{1}{z_a (\beta_1)}\,e^{-(\beta_1+i\mu)\hbar\omega (s+m-q)}\,\underbrace{\sum_{j=0}^\infty \frac{(s+m-q+j)!}{j!}\,e^{-(\beta_1+i\mu)\hbar\omega j}}_{\frac{(s+m-q)!}{[1-e^{-(\beta_1+i\mu)\hbar\omega}]^{s+m-q+1}}},\nn\\
 =&& \frac{1}{z_a (\beta_1)}\,z_a (\beta_1+i\mu)\,\frac{(s+m-q)!}{\Big(e^{(\beta_1+i\mu)\hbar\omega}-1\Big)^{s+m-q}}.\nn
\eeq
Therefore,

$$ G(\mu,t)= \sum_{m=0}^\infty \frac{e^{i\mu\hbar\omega m}}{m!}\sum_{s=0}^\infty \frac{(-1)^s}{s!}\sum_{q=0}^{s+m}\frac{[(s+m)!]^2\,|f(t)|^{2(s+m-q)}\big(|h(t)|^2 \bar{n}_b\big)^q}{q! (s+m-q)!}$$
\beq\label{s25}
\,\times\frac{z_a (\beta_1+i\mu)}{z_a (\beta_1)\,\Big(e^{(\beta_1+i\mu)\hbar\omega}-1\Big)^{s+m-q}},\nn
\eeq
\beq
&=& \frac{z_a (\beta_1+i\mu)}{z_a (\beta_1)}\,\sum_{m=0}^\infty \frac{e^{i\mu\hbar\omega m}}{m!}\sum_{s=0}^\infty \frac{(-1)^s (s+m)!}{s!}\nn\\
&&\times\,\Big(\frac{|f(t)|^2+|h(t)|^2 \bar{n}_b\,(e^{(\beta_1+i\mu)\hbar\omega}-1)}{e^{(\beta_1+i\mu)\hbar\omega}-1}\Big)^{s+m},\nn\\
&=& \frac{z_a (\beta_1+i\mu)}{z_a (\beta_1)}\,\sum_{m=0}^\infty \frac{\big[e^{i\mu\hbar\omega}\,\Big(\frac{|f(t)|^2}{e^{(\beta_1+i\mu)\hbar\omega}-1}+|h(t)|^2 \bar{n}_b\Big)\big]^m}{m!}\nn\\
&&\times\,\frac{m!}{\Big(\frac{|f(t)|^2}{e^{(\beta_1+i\mu)\hbar\omega}-1}+|h(t)|^2 \bar{n}_b+1\Big)^{m+1}},\nn
\eeq
$$
=\frac{z_a (\beta_1 +i\mu)}{z_a (\beta_1)}\frac{1}{\Big[\frac{|f(t)|^2}{e^{(\beta_1+i\mu)\hbar\omega}-1}+|h(t)|^2 \bar{n}_b+1-e^{i\mu\hbar\omega}\Big(\frac{|f(t)|^2}{e^{(\beta_1+i\mu)\hbar\omega}-1}+|h(t)|^2 \bar{n}_b\Big)\Big]},
$$
$$
=\frac{\big(e^{\beta_1\hbar\omega}-1\big)\,e^{i\mu\hbar\omega}}{e^{i\mu\hbar\omega}\big[(1+|h(t)|^2 \bar{n}_b)e^{\beta_1\hbar\omega}-|h(t)|^2 \bar{n}_b e^{(\beta_1+i\mu)\hbar\omega}+|h(t)|^2 (\bar{n}_b+1)-1\big]-|h(t)|^2 (\bar{n}_b+1)}.
$$
\subsection{Derivation of (\ref{two2})}
\noindent Hamiltonian (18) can be rewritten in matrix form as
\be\label{s26}
\hat{H} (t)=\hbar
\left(
\begin{array}{ccc}
                 \hat{a}^\dag & \hat{b}^\dag & \hat{c}^\dag \\
                 \end{array}
                  \right)
\underbrace{\left(
                  \begin{array}{ccc}
                     \omega & g(t) & g(t) \\
                       g(t) & \omega & 0 \\
                       g(t) & 0 & \omega \\
                  \end{array}
            \right)}_{\Lambda(t)}\left(
                                 \begin{array}{c}
                                   \hat{a} \\
                                   \hat{b} \\
                                   \hat{c} \\
                                 \end{array}
                                \right).\nn
\ee
The time-dependent matrix $\Lambda(t)$ exhibits commutativity at distinct times, denoted by $[\Lambda(t),\Lambda(t')] = 0$, with eigenvalues $\lambda_1=\omega$, $\lambda_2=\omega-\sqrt{2}\,g(t)$, and $\lambda_3=\omega+\sqrt{2}\,g(t)$. The associated renormalized eigenvectors are defined as
$$
\ket{\lambda_1}=\left(
                  \begin{array}{c}
                    0 \\
                    \sqrt{2}/2 \\
                    -\sqrt{2}/2 \\
                  \end{array}
                \right),\,\,\,\,\,\ket{\lambda_2}=\left(
                                                    \begin{array}{c}
                                                 -\sqrt{2}/2 \\
                                                      1/2 \\
                                                      1/2 \\
                                                    \end{array}
                                                  \right),\,\,\,\,\,\ket{\lambda_3}=\left(
                                                    \begin{array}{c}
                                               \sqrt{2}/2 \\
                                                      1/2 \\
                                                      1/2 \\
                                                    \end{array}
                                                  \right).\nn
$$
The orthogonal matrix $V$ generated from these eigenvectors is represented as
\be\label{s28}
V=\left(
    \begin{array}{ccc}
      0 & -\sqrt{2}/2 & \sqrt{2}/2 \\
      \sqrt{2}/2 & 1/2 & 1/2 \\
      -\sqrt{2}/2 & 1/2 & 1/2 \\
    \end{array}
  \right).\nn
\ee
We utilize the orthogonal matrix $V$ to diagonalize the Hamiltonian $\hat{H}$ and introduce new operators $\hat{A}$, $\hat{B}$, and $\hat{C}$ as
\be\label{s29}
\left(
  \begin{array}{c}
    \hat{A} \\
    \hat{B} \\
    \hat{C} \\
  \end{array}
\right)=\left(
    \begin{array}{ccc}
      0 & -\sqrt{2}/2 & \sqrt{2}/2 \\
      \sqrt{2}/2 & 1/2 & 1/2 \\
      -\sqrt{2}/2 & 1/2 & 1/2 \\
    \end{array}
  \right)\left(
           \begin{array}{c}
             \hat{a} \\
             \hat{b} \\
             \hat{c} \\
           \end{array}
         \right).
\ee
The Hamiltonian in terms of the new operators is given by
\be\label{s30}
\hat{H}(t)=\hbar\omega\,\hat{A}^\dag \hat{A}+\hbar(\omega-\sqrt{2}\,g(t))\,\hat{B}^\dag \hat{B}+\hbar(\omega+\sqrt{2}\,g(t))\,\hat{C}^\dag \hat{C}.\nn
\ee
From the Heisenberg equations of motion for the new operators, we deduce
\beq\label{s31}
\hat{A}(t) &=& e^{-i\omega t}\,\hat{A} (0),\,\,\,\,\hat{B}(t)=e^{-i(\omega t-\sqrt{2}\,G(t))}\,\hat{B} (0),\nn\\
\hat{C}(t) &=&e^{-i(\omega t+\sqrt{2}\,G(t))}\,\hat{C} (0),\nn
\eeq
where $G(t)=\int_0^t dt'\,g(t')$. By inverting the matrix equation (\ref{s29}), we obtain
$$
\hat{a} (t)=e^{-i\omega t}\,\cos(\sqrt{2}G(t))\,\hat{a} (0)+\frac{i\sqrt{2}}{2}\,e^{-i\omega t}\,\sin(\sqrt{2}G(t))\,(\hat{b}(0)+\hat{c}(0)).
$$
\subsection{Derivation of (\ref{TL7})}
\noindent By substituting the state
$$
\ket{\psi(t)}=C_{++}\ket{+}_1\ket{+}_2+C_{+-}\ket{+}_1\ket{-}_2+C_{-+}\ket{-}_1\ket{+}_2+C_{--}\ket{-}_1\ket{-}_2,
$$
into the Schr\"{o}dinger equation with Hamiltonian (\ref{TL1}), we obtain
\beq\label{s34}
&& C_{++} (t)=e^{-i\omega t}\,C_{++} (0),\nn\\
&& C_{--} (t)=e^{i\omega t}\,C_{--} (0),\nn\\
&& \dot{C}_{+-} (t)=-i g(t)\,C_{-+} (t),\nn\\
&& \dot{C}_{-+} (t)=-i g(t)\,C_{+-} (t).
\eeq
By employing the new variables
\beq\label{s35}
&& C_1 (t)=\frac{1}{\sqrt{2}}(C_{+-} (t)+C_{-+} (t)),\nn\\
&& C_2 (t)=\frac{1}{\sqrt{2}}(C_{+-} (t)-C_{-+} (t)),
\eeq
the coupled equations for the coefficients $C_{+-}$ and $C_{-+}$ decouple, and we obtain
\beq\label{s36}
&& C_{+-} (t)=\cos(G(t))\,C_{+-} (0)-i\sin(G(t))\,C_{-+} (0),\nn\\
&& C_{-+} (t)=\cos(G(t))\,C_{-+} (0)-i\sin(G(t))\,C_{+-} (0).
\eeq
In the standard basis, the Schr\"{o}dinger equation can be expressed in the matrix form as
$$
\footnotesize{\left(
  \begin{array}{c}
    C_{++} (t) \\
    C_{+-} (t) \\
    C_{-+} (t) \\
    C_{--} (t) \\
  \end{array}
\right)=\underbrace{\left(
          \begin{array}{cccc}
            e^{-i\omega_0 t} & 0 & 0 & 0 \\
            0 & \cos (G(t)) & -i\sin(G(t)) & 0 \\
            0 & -i\sin(G(t)) & \cos (G(t)) & 0 \\
            0 & 0 & 0 & e^{i\omega_0 t} \\
          \end{array}
        \right)}_{\hat{W}(t)}\left(
  \begin{array}{c}
    C_{++} (0) \\
    C_{+-} (0) \\
    C_{-+} (0) \\
    C_{--} (0) \\
  \end{array}
\right)},
$$
where the matrix $\hat{W}(t)$ ia the unitary evolution matrix ($\hat{W}(t)\,\hat{W}^\dag (t)=\hat{W}^\dag (t)\,\hat{W}(t)=i_d$). Suppose the two-level atom is initially prepared in an arbitrary state and the bath is held in a thermal state, then the initial state is expressed as
$$
\hat{\rho} (0)=\left(
                 \begin{array}{cc}
                   a & \bar{c} \\
                   c & b \\
                 \end{array}
               \right)\otimes \left(
                                \begin{array}{cc}
                                  P_e & 0 \\
                                  0 & P_g \\
                                \end{array}
                              \right)=\left(
                                        \begin{array}{cccc}
                                          a P_e & 0 & \bar{c} P_e & 0 \\
                                          0 & a P_g & 0 & \bar{c} P_g \\
                                          c P_e & 0 & b P_e & 0 \\
                                          0 & c P-g & 0 & b P_g \\
                                        \end{array}
                                      \right).
$$
The total density matrix at an arbitrary time $t$ is given by
$$\hat{\rho} (t)= \hat{W}(t)\,\hat{\rho} (0)\,\hat{W}^\dag (t),$$
$$
    =\footnotesize{\left(
          \begin{array}{cccc}
            e^{-i\omega_0 t} & 0 & 0 & 0 \\
            0 & \cos G(t) & -i\sin G(t) & 0 \\
            0 & -i\sin G(t) & \cos G(t) & 0 \\
            0 & 0 & 0 & e^{i\omega_0 t} \\
          \end{array}
        \right)\left(
                                        \begin{array}{cccc}
                                          a P_e & 0 & \bar{c} P_e & 0 \\
                                          0 & a P_g & 0 & \bar{c} P_g \\
                                          c P_e & 0 & b P_e & 0 \\
                                          0 & c P-g & 0 & b P_g \\
                                        \end{array}
                                      \right)}
$$
$$
\times
                                   \footnotesize{\left(
          \begin{array}{cccc}
            e^{i\omega_0 t} & 0 & 0 & 0 \\
            0 & \cos G(t) & i\sin G(t) & 0 \\
            0 & i\sin G(t) & \cos G(t) & 0 \\
            0 & 0 & 0 & e^{-i\omega_0 t} \\
          \end{array}
        \right)},
$$
therefore,
$$
\hat{\rho} (t) =\tiny{\left(
      \begin{array}{cccc}
        a P_e & i\sin G e^{-i\omega_0 t} \bar{c} P_e & \cos G e^{-i\omega_0 t} \bar{c} P_e & 0 \\
        -i\sin G e^{i\omega_0 t} c P_e & a P_g \cos^2 G+b P_e \sin^2 G & i\sin G\cos G (a P_g-b P_e) & \cos G e^{-i\omega_0 t} \bar{c} P_g \\
        \cos G e^{i\omega_0 t} c P_e & -i\sin G\cos G (a P_g-b P_e) & a P_g \sin^2 G+b P_e \cos^2 G & -i\sin G e^{-i\omega_0 t} \bar{c}\,P_g \\
        0 & \cos G e^{i\omega_0 t} c P_g & i\sin G e^{i\omega_0 t} c P_g & b P_g \\
      \end{array}
    \right)}.
$$
By tracing out the bath degrees of freedom, we find the reduced density matrix as
$$
\hat{\rho}_a (t)=\left(
                   \begin{array}{cc}
                     \phi(t) & \bar{c} e^{-i\omega_0 t}\cos(G(t)) \\
                     c e^{i\omega_0 t}\cos(G(t)) & \psi(t) \\
                   \end{array}
                 \right),
$$
where
\beq
&& \phi(t)=a\cos^2 (G(t))+P_e\sin^2 (G(t)),\nn\\
&& \psi(t)=b\cos^2 (G(t))+P_g\sin^2 (G(t)).\nn
\eeq
\subsection{Derivation of (\ref{dephase3})}
\noindent Let the time-dependent operator $\hat{B} (t)$ at the ordered discrete times $t_1, t_2, \cdots, t_{n-1}, t_{n}$ be denoted by $\hat{B}_1, \hat{B}_1,\cdots,\hat{B}_n$, respectively. Suppose that for any $t$ and $t'$, the commutator $[\hat{B} (t),\hat{B} (t')]$ be a $c$-number. Then the generalized form of the Baker-Campbell-Hausdorff formula is expressed as
\be\label{s41}
e^{\sum\limits_{i=1}^n \hat{B}_i}=e^{\hat{B}_1}\,e^{\hat{B}_2}\cdots e^{\hat{B}_{n-1}}\,e^{\hat{B}_n}\,e^{-\sum\limits_{i<j}\frac{1}{2}[\hat{B}_i, \hat{B}_j]}.\nn
\ee
The evolution operator corresponding to the Hamiltonian (\ref{dephase1}) in the interaction picture is given by
\be\label{sUint}
\hat{U}_I (t)= T\,e^{-\frac{i}{\hbar}\int\limits_0^t dt'\,\hat{H}_I (t')},\nn
\ee
where $T$ represents the time-ordering operator. Since $[\hat{H}_I (t),\hat{H}_I (t')]=0$, we can utilize (\ref{s41}) by setting $\hat{B}_i\equiv \hat{H}_I (i\Delta t)\Delta t$, and rewrite the propagator as
\beq\label{s43}
&&\hat{U}_I (t)=T \lim_{\{N\to\infty, \Delta t\to 0, N\Delta t=t\}}\,e^{\sum\limits_{n=1}^N -\frac{i}{\hbar}\,\hat{H}_I (n\Delta t)\,\Delta t},\nn\\
&&             = T \underbrace{\prod_{n=1}^N e^{-\frac{i}{\hbar}\,\hat{H}_I (n\Delta t)\,\Delta t}}_{\mbox{time-ordered}}\,e^{\sum\limits_{i<j}\frac{1}{2}[\hat{H}_I (i\Delta t), \hat{H}_I (j\Delta t)]\frac{(\Delta t)^2}{\hbar^2}},\nn\\
&&             = \prod_{n=1}^N e^{-\frac{i}{\hbar}\,\hat{H}_I (n\Delta t)\,\Delta t}\,e^{\sum\limits_{i<j}\frac{1}{2}[\hat{H}_I (i\Delta t), \hat{H}_I (j\Delta t)]\frac{(\Delta t)^2}{\hbar^2}},\nn\\
&&             = e^{-\frac{i}{\hbar}\int\limits_0^t dt'\,\hat{H}_I (t')}=e^{\sigma_z\otimes\big[-i\int\limits_{0}^t dt'\,g(t')e^{-i\omega t'}\,\hat{b}-i\int\limits_{0}^t dt'\,g(t')e^{i\omega t'}\,\hat{b}^\dag\big]},\nn\\
&&             = e^{\sigma_z\otimes \hat{A}(t)},\nn
\eeq
where we defined
\beq\label{s44}
&& \hat{A}(t)=\xi(t)\hat{b}-\bar{\xi}(t)\hat{b}^\dag,\nn\\
&& \xi(t)=-i\int_0^t dt'\,g(t') e^{-i\omega t'}.\nn
\eeq
Expanding $e^{\sigma_z\otimes \hat{A}(t)}$, we obtain
\beq\label{s45}
\hat{U}_I (t)&=&e^{\sigma_z\otimes \hat{A}(t)}= i_d+\sigma_z\otimes\hat{A}+\frac{1}{2!}\hat{A}^2+\frac{1}{3!}\sigma_z\otimes\hat{A}^3+\frac{1}{4!}\hat{A}^4+\cdots,\nn\\
             &=& i_d\otimes\cosh(\hat{A})+\sigma_z\otimes\sinh(\hat{A}).\nn
\eeq
\subsection{Derivation of (\ref{dephase5})}
\noindent For the initial state
\[
\hat{\rho} (0)=\left(
                 \begin{array}{cc}
                   a & \bar{c} \\
                   c & b \\
                 \end{array}
               \right)\otimes\frac{-\beta\hbar\omega\hat{b}^\dag\hat{b}}{z_b},
\]
by making use of the evolution operator in Schr\"{o}dinger picture $\hat{U} (t)=\hat{U}_0 (t)\hat{U}_I (t)$, we find the reduced density matrix as follows
\begin{eqnarray}
\hat{\rho}_s (t) &=& tr_a\big(\hat{U}(t)\,\hat{\rho}(0)\,\hat{U}^\dag (t)\big),\nn\\
                 &=& \left(
                     \begin{array}{cc}
                       a & \bar{c}e^{-i\omega_0 t} \\
                       ce^{i\omega_0 t} & b \\
                     \end{array}
                   \right)\,tr_a \bigg(\cosh^2 (\hat{A})\,\frac{e^{-\beta\hbar\omega\hat{b}^\dag \hat{b}}}{z_b}\bigg)\nn\\
&&\,\,\,\,                   -\left(
                     \begin{array}{cc}
                       a & -\bar{c}e^{-i\omega_0 t} \\
                       ce^{i\omega_0 t} & -b \\
                     \end{array}
                   \right)\,tr_a \bigg(\sinh(\hat{A})\cosh(\hat{A})\frac{e^{-\beta\hbar\omega\hat{b}^\dag \hat{b}}}{z_b}\bigg)\nn\\
&&\,\,\,\,                  + \left(
                     \begin{array}{cc}
                       a & \bar{c}e^{-i\omega_0 t} \\
                       -ce^{i\omega_0 t} & -b \\
                     \end{array}
                   \right)\,tr_a \bigg(\sinh(\hat{A})\cosh(\hat{A})\frac{e^{-\beta\hbar\omega\hat{b}^\dag \hat{b}}}{z_b}\bigg)\nn\\
&& \,\,\,\,                  -\left(
                     \begin{array}{cc}
                       a & -\bar{c}e^{-i\omega_0 t} \\
                       -ce^{i\omega_0 t} & b \\
                     \end{array}
                   \right)
                   \,tr_a \bigg(\sinh^2 (\hat{A})\,\frac{e^{-\beta\hbar\omega\hat{b}^\dag \hat{b}}}{z_b}\bigg).\nn
\end{eqnarray}
To proceed, we use
\beq\label{s48}
tr_a \bigg(e^{2\hat{A}}\,\frac{e^{-\beta\hbar\omega_0\hat{b}^\dag \hat{b}}}{z_b}\bigg)&=& \frac{1}{z_b}\,tr_a\bigg[e^{2\xi(t)\hat{b}-2\bar{\xi}(t)\hat{b}^\dag}\,e^{-\beta\hbar\omega\hat{b}^\dag\hat{b}}\bigg],\nn\\
                                                                                  &=& \frac{1}{z_b}\sum_{n=0}^\infty e^{-n\beta\hbar\omega}\bra{n}e^{-2\bar{\xi}(t)\hat{b}^\dag}e^{2\xi(t)\hat{b}}e^{-2|\xi(t)|^2}\ket{n},\nn\\
                                                                                  &=& \frac{e^{-2|\xi(t)|^2}}{z_b}\sum_{n=0}^\infty e^{-n\beta\hbar\omega}
                                                                                  \underbrace{\bra{n}e^{-2\bar{\xi}(t)\hat{b}^\dag}e^{2\xi(t)\hat{b}}\ket{n}}_{L_n (4|\xi(t)|^2)},\nn\\
                                                                                  &=& \frac{e^{-2|\xi(t)|^2}}{z_b}\,\frac{e^{\beta\hbar\omega_0}\,e^{\frac{4|\xi(t)|^2}{1-e^{\beta\hbar\omega}}}}{e^{\beta\hbar\omega}-1},\nn\\
                                                                                  &=& e^{-2|\xi(t)|^2}\,e^{\frac{4|\xi(t)|^2}{1-e^{\beta\hbar\omega}}},\nn\\
                                                                                  &=& e^{-2|\xi(t)|^2\coth(\beta\hbar\omega/2)}.\nn
\eeq
where we made use of equations (\ref{s18}) and (\ref{s19}). Also, from $\hat{A}^\dag=-\hat{A}$, we have
\beq\label{s49}
tr_a \bigg(e^{-2\hat{A}}\,\frac{e^{-\beta\hbar\omega\hat{b}^\dag \hat{b}}}{z_b}\bigg)=tr_a \bigg(e^{2\hat{A}}\,\frac{e^{-\beta\hbar\omega\hat{b}^\dag \hat{b}}}{z_b}\bigg)=e^{-2|\xi(t)|^2\coth(\beta\hbar\omega/2)}.\nn
\eeq
Now, we easily obtain
\beq\label{s50}
&& tr_a \bigg(\cosh^2 (\hat{A})\,\frac{e^{-\beta\hbar\omega\hat{b}^\dag \hat{b}}}{z_b}\bigg)=tr_a \bigg(\frac{(e^{\hat{A}}+e^{-\hat{A}})^2}{4}\,\frac{e^{-\beta\hbar\omega\hat{b}^\dag \hat{b}}}{z_b}\bigg),\nn\\
                                                                                          &&= \frac{1}{2}\,\bigg(e^{-2|\xi(t)|^2\coth(\beta\hbar\omega/2)}+1\bigg),\nn\\
&& tr_a \bigg(\sinh^2 (\hat{A})\,\frac{e^{-\beta\hbar\omega\hat{b}^\dag \hat{b}}}{z_b}\bigg)=tr_a \bigg(\frac{(e^{\hat{A}}-e^{-\hat{A}})^2}{4}\,\frac{e^{-\beta\hbar\omega\hat{b}^\dag \hat{b}}}{z_b}\bigg),\nn\\
                                                                                          &&= \frac{1}{2}\,\bigg(e^{-2|\xi(t)|^2\coth(\beta\hbar\omega/2)}-1\bigg),\nn
\eeq
$$tr_a \bigg(\sinh(\hat{A})\cosh(\hat{A})\,\frac{e^{-\beta\hbar\omega\hat{b}^\dag \hat{b}}}{z_b}\bigg) = \frac{1}{4}\,\bigg((e^{\hat{A}}-e^{-\hat{A}})(e^{\hat{A}}+e^{-\hat{A}})\,\frac{e^{-\beta\hbar\omega\hat{b}^\dag \hat{b}}}{z_b}\bigg)=0.$$                                                                                        Therefore,
$$
\hat{\rho}_s (t)=\left(
                   \begin{array}{cc}
                     a & e^{-2|\xi(t)|^2\coth(\beta\hbar\omega/2)}e^{-i\omega_0 t}\,\bar{c} \\
                     e^{-2|\xi(t)|^2\coth(\beta\hbar\omega/2)}e^{i\omega_0 t}\,c & b \\
                   \end{array}
                 \right).\nn
$$
\newpage


\end{document}